\def\spose#1{\hbox to 0pt{#1\hss}}
\def\simlt{\mathrel{\spose{\lower 3pt\hbox{$\mathchar"218$}}
     \raise 2.0pt\hbox{$\mathchar"13C$}}}
\def\simgt{\mathrel{\spose{\lower 3pt\hbox{$\mathchar"218$}}
     \raise 2.0pt\hbox{$\mathchar"13E$}}}
\def\simpropto{\mathrel{\spose{\lower 3pt\hbox{$\mathchar"218$}}
     \raise 2.0pt\hbox{$\propto$}}}

\documentclass[twocolumn,aps,prd,nofootinbib,showpacs]{revtex4}
\usepackage{amsmath,graphicx,bm}
\begin{document}
\title{A Method for 21\,cm Power Spectrum Estimation in the Presence of Foregrounds}

\author{Adrian Liu}
\email{acliu@mit.edu}
\author{Max Tegmark}
\affiliation{Dept. of Physics and MIT Kavli Institute, Massachusetts Institute of Technology, Cambridge, MA 02139, USA}
\date{\today}


\begin{abstract}
$21\,\textrm{cm}$ tomography promises to be a powerful tool for estimating cosmological parameters, constraining the epoch of reionization, and probing the so-called dark ages.  However, realizing this promise will require the extraction of a cosmological power spectrum from beneath overwhelmingly large sources of foreground contamination.  In this paper, we develop a unified matrix-based framework for foreground subtraction and power spectrum estimation, which allows us to quantify the errors and biases that arise in the power spectrum as a result of foreground subtraction.  We find that existing line-of-sight foreground subtraction proposals can lead to substantial mode-mixing as well as residual noise and foreground biases, whereas our proposed inverse variance foreground subtraction eliminates noise and foreground biases, gives smaller error bars, and produces less correlated measurements of the power spectrum.  We also numerically confirm the intuitive belief in the literature that $21\,\textrm{cm}$ foreground subtraction is best done using frequency rather than angular information.
\end{abstract}

\maketitle

\section{Introduction}
In recent years, $21\,\textrm{cm}$ tomography has been hailed as a cosmological probe with the potential to overtake Cosmic Microwave Background (CMB) measurements as the most promising tool for understanding our Universe.  Much of this is due to the sheer information content available --- as an optically thin transition of the most abundant element in the Universe, the redshifted $21\,\textrm{cm}$ hyperfine transition line allows one to probe a much larger volume of our Universe than ever before.  This is projected to lead to better constraints on cosmological parameters related to inflation, dark matter, dark energy, and neutrinos \citep{Matt3,Santos2, juddjackiemiguel1,Yi,Whitepaper2,wyithe2008,ChangDE,miguelreview}.  Moreover, $21\,\textrm{cm}$ tomography may be our only direct probe of the Epoch of Reionization (EoR) as well as the preceding cosmological dark ages \citep{Rees, Tozzi2, Tozzi, Iliev,furlanetto1,Loeb1,furlanetto2, Barkana1,Mack,Whitepaper1}.

Despite its great potential, several difficulties must be overcome for $21\,\textrm{cm}$ tomography to become a useful scientific tool.  That the cosmological signal is expected to be on the order of $10\,\textrm{mK}$ in brightness temperature means that sensitivity requirements for a mere detection will be high, especially given that foreground contamination of various origins (such as unresolved extragalactic point sources, resolved point sources, and Galactic synchrotron emission) are expected to contribute a total contaminating effect in the range of hundreds of Kelvins \citep{Angelica}.  As a result, the first-generation of arrays currently being tested for cosmological $21\,\textrm{cm}$ measurements (such as PAPER \citep{PAPER}, LOFAR \citep{LOFARinstrument}, MWA \citep{MWAdesign}, CRT \citep{CRT}, GMRT \citep{GMRT}, and the Omniscope \citep{FFTT2}) are not expected to be capable of producing images of the small ionized regions that formed around the first luminous sources during the EoR.  Instead, any cosmological results coming from these arrays will likely be based on large scale \emph{statistical} measures.  One example would be the so-called \emph{global step} measurements that seek to measure the evolution of the average $21\,\textrm{cm}$ signal of the whole sky \citep{EDGES}.  Going beyond this, one may seek to detect the spatial \emph{fluctuations} of the signal and to estimate the power spectrum.  The power spectrum, in particular, has thus far been shown to be the most promising 
way to leverage $21\,\textrm{cm}$ tomography for precision cosmology, potentially placing extremely tight constraints on parameters such as the neutrino mass, the dark energy equation of state, and spatial curvature \citep{Yi,Matt3,pritchardneutrinos}.

Power spectrum estimation in $21\,\textrm{cm}$ tomography has similarities to and differences from power spectrum estimation for both the CMB and galaxy surveys.  As it is for the CMB, foreground contamination is a serious concern, especially since the foregrounds are so strong at the lower frequencies ($\sim\!\!100-200\,\textrm{MHz}$) that EoR experiments target.  Unlike the CMB, however, $21\,\textrm{cm}$ tomography probes a three-dimensional volume (just like with galaxy surveys), so the final quantity of interest is not an angular power spectrum but a three-dimensional power spectrum.  Ultimately, one wishes to obtain the spherically averaged matter power spectrum as a function of redshift $P(k,z)$, because one expects the cosmological power to be isotropic.  However, it is often preferable to first form a cylindrically averaged power spectrum $P(k_\perp, k_\parallel; z)$, since the isotropy may be destroyed by redshift-space distortions \citep{Barkana2}, the Alcock-Pacyznski effect \citep{AliAP, NusserAP,BarkanaAP}, as well as residual foregrounds that arise from imperfect foreground removal.

The problem of foreground removal in $21\,\textrm{cm}$ tomography has been extensively studied in the literature.  Early foreground cleaning proposals focused on using the angular structure of the foregrounds \citep{dimatteo1,dimatteo2,Oh,Santos,ZFH}, while recent studies have suggested that a line-of-sight (LOS) spectral subtraction method may be the most promising approach \citep{xiaomin,nusserforegrounds,Judd08,paper1,LOFAR,Harker,paper2,LOFAR2}.  Though the detailed algorithmic implementations of the LOS method vary amongst these studies, the core idea is always to take advantage of the smooth spectral behavior of foregrounds to separate them from any cosmological signals, which are expected to fluctuate rapidly with frequency.  Simulations of this approach have confirmed it as a successful foreground cleaning technique in the sense that the post-subtraction foreground residuals can be suppressed to a level smaller than the expected amplitude of the cosmological signal.  Thus, foreground contamination is unlikely to be an insurmountable obstacle to the initial \emph{detection} of the $21\,\textrm{cm}$ cosmological signal.

What remains unclear, however, is whether LOS subtraction techniques adversely affect the quality of one's final estimate of the cosmological power spectrum.  For instance, although spectrally smooth components of the cosmological signal are expected to be small, if any are present they will be inadvertently subtracted off with the foregrounds when LOS cleaning is performed.  LOS methods are thus not \emph{lossless}, in the sense that their destruction of cosmological information leads to a degradation of error bars on the final power spectra (for a more rigorous definition of information loss, see \citep{Maxpowerspeclossless} or Section \ref{formalism} of this paper for a quick review).  Foreground subtraction may also lead to residual systematic noise and foreground \emph{biases} in estimates of the power spectrum, a fact that was explored numerically using simulations in \citep{Judd08,LOFAR2,PetrovicOh}.   Finally, foreground subtraction along the LOS may lead to correlated errors in power spectrum measurements, which can limit their usefulness for estimating cosmological parameters.

The goal of this paper is to adapt the existing mathematical formalism for power spectrum estimation from the CMB and galaxy survey literature in a way that not only respects the unique geometrical properties of $21\,\textrm{cm}$ tomography experiments, but also allows us to robustly deal with the issue of foreground contamination.  This will permit a more complete analysis of the errors involved in estimating $21\,\textrm{cm}$ power spectra, in a way that automatically incorporates the effect that foreground removal has on the final results.  In particular, we will be able to quantify not just the errors on the power spectrum itself (the ``vertical" error bars), but also the correlations between the parts of the power spectrum (the ``horizontal" error bars, or equivalently the power spectrum window functions\footnote{These are not to be confused with the instrumental window functions that are conventionally defined in radio astronomy to quantify the effect of an instrument's beam and noise properties (and as used, for instance, in related $21\,\textrm{cm}$ power spectrum papers such as \citep{MiguelJackie1} and \citep{Judd06}).  In this paper, the ``window function" can be thought of as a convolution kernel that produces the measured power spectrum when applied to the true power spectrum.  Please see Section \ref{formalism} for a rigorous mathematical definition.}).  In addition, we will quantify the noise and foreground biases that are introduced by foreground subtraction and power spectrum estimation so that such biases can be systematically removed.  We accomplish all this by considering the foregrounds not as an additional signal to be removed, but rather as a form of correlated noise.  Doing so in our formalism allows us to build on the numerical simulation work of \citep{Judd08}, \citep{LOFAR2}, and \citep{PetrovicOh} to \emph{analytically} show how LOS foreground subtraction methods are lossy and leave residual noise and foreground biases.  The formalism also naturally suggests an alternative method for foreground subtraction that leads to smaller error bars and eliminates the residual biases.

The rest of this paper is organized as follows.  In Section \ref{formalism}, we introduce the mathematical notation and formalism that we use for power spectrum estimation.  Existing LOS methods are cast in this language, and in addition we introduce our alternative foreground subtraction scheme.  In Section \ref{results}, we test our methods on a specific foreground model, which we describe in Section \ref{foregroundsandnoise}.  Provided that the foregrounds satisfy certain generic criteria (such as having a smooth frequency dependence), the details of the foreground model should have no effect on our qualitative conclusions.  In Section \ref{toy}, we show how these qualitative conclusions can be understood through a simple toy model.  We summarize our conclusions and discuss broader implications in Section \ref{conc}.
\section{The Mathematical Framework}
\label{formalism}
In this section we review the power spectrum estimation formalism introduced by \citep{Maxpowerspeclossless}, and adapt it for $21\,\textrm{cm}$ tomography.  Readers interested in the mathematical details of power spectrum estimation are encouraged to peruse \citep{BJK,Maxpowerspeclossless,Maxgalaxysurvey1,Maxgalaxysurvey2}.  Our development builds on the quadratic methods presented in \citep{Maxpowerspeclossless,Maxgalaxysurvey1}.  The goal is to derive new expressions appropriate to $21\,\textrm{cm}$ power spectrum estimation (such as Equation \ref{cylindQ} and the expressions in Section \ref{LOS}) as well as to systematically place different power spectrum estimation methods in a common matrix-based framework.
\subsection{Quadratic Estimators}
\label{quadest}
Our objective is to use the measured $21\,\textrm{cm}$ brightness temperature distribution $T_{b} (\mathbf{r})$ to estimate a cylindrically symmetric power spectrum\footnote{Note that in this paper, we are not examining the steps required to estimate the \emph{matter} power spectrum.  Instead, we are focussing on estimating the \emph{temperature} power spectrum $P_T (k_{\perp},k_{\parallel})$, which is an important first step for finding the matter power spectrum.  For details on how one might extract a matter power spectrum from a temperature power spectrum, see for example \citep{Barkana2,Yi}.} $P_T (k_{\perp},k_{\parallel})$ which is conventionally defined by the equation
\begin{equation}
\label{Pdef}
\langle \widehat{T}_b (\mathbf{k})^* \widehat{T}_b (\mathbf{k}^{\prime}) \rangle = ( 2\pi)^3 \delta(\mathbf{k}-\mathbf{k}^{\prime}) P_T (k_{\perp},k_{\parallel}),
\end{equation}
where $\widehat{T}_b$ is the three-dimensional spatial Fourier transform of $T_{b}$, and $\delta$ is the Dirac delta function.  As written, this definition contains the \emph{continuous} functions $T_b$ and $P_T$.  However, any practical numerical scheme for estimating the power spectrum must necessarily be discrete.  The simplest way to do this is to divide the measured spatial distribution of brightness temperature into discrete voxels, so that $T_b$ takes the form of a data vector $\mathbf{x}$ that is essentially a list of the brightness temperatures measured at various points on a three-dimensional grid.  
Note that in forming this vector, one should \emph{not} use the full radial extent of the data that are available in a typical $21\,\textrm{cm}$ tomography experiment.  Instead, one should split the full data into \emph{multiple} $\mathbf{x}$ vectors, each of which spans only a very narrow range in redshift so that the evolution of the cosmological signal is negligible.  The analysis that we describe in this paper should then be performed \emph{separately} on each vector to produce a number of power spectra at different redshifts.  Failure to do so would violate the assumption of translational invariance that was implicit when we defined the power spectrum using Equation \ref{Pdef}.

To discretize the power spectrum, we parametrize it as a piecewise \emph{constant} function\footnote{In a practical application of the methods described in this paper, it is often preferable to instead parametrize as a piecewise constant function the \emph{ratio} of the power spectrum to a prior.  As shown in \citep{HamTeg}, doing so tends to give better behaved window functions (Equation \ref{wind}).  Here for simplicity we employ a white (\emph{i.e.} constant) prior simply because there are as yet no observational constraints on the form of the $21\,\textrm{cm}$ power spectrum.}, such that $P_T (k_{\perp},k_{\parallel}) = p_{ab}$ for $k^{\perp}_a \le k^{\perp} < k^{\perp}_{a+1}$ and $k^{\parallel}_b \le k^{\parallel} < k^{\parallel}_{b+1}$.  If the index $a$ runs over $M$ different values and the index $b$ runs over $N$ values, the power spectrum can then be stored in an $MN$-dimensional vector $p_{\alpha}$, where index pairs $(a,b)$ have been folded into a single index $\alpha$.  Parameterizing the power spectrum in this way (where each component $p_{\alpha}$ is referred to as the \emph{band power} of the band $\alpha$) represents no significant loss of information as long as the widths of the $k$ bins are small compared to the physical scales over which $P_T$ varies appreciably \citep{Maxgalaxysurvey1}.

A \emph{quadratic method} for estimating our discretized power spectrum is one where the estimator of $p_\alpha$ takes the quadratic form
\begin{equation}
\label{quadmethod}
\widehat{p}_\alpha = \mathbf{(x-m)}^t \mathbf{E}^{\alpha} \mathbf{(x-m)} - b_\alpha
\end{equation}
for some family of symmetric matrices $\mathbf{E}^{\alpha}$ and constants $b_\alpha$ (one for each $k$-region in our piecewise constant discretized power spectrum).  The vector $\mathbf{m}$ is defined as the \emph{ensemble average} over different random realizations of the random variable data vector $\mathbf{x}$, \emph{i.e.} $\mathbf{m} \equiv \langle \mathbf{x} \rangle$.  The hat ( $\widehat{}$ ) denotes the fact that what we have here is an \emph{estimate} of the true power spectrum $p_\alpha$ from the data.  The matrix $\mathbf{E}^{\alpha}$ encodes the Fourier transforms, binning, and --- crucially --- the weighting and foreground subtraction steps required in going from the calibrated data vector $\mathbf{x}$ to the corresponding estimate of the power spectrum $p_{\alpha}$.  For example, \emph{if one chooses to forgo any form of foreground subtraction} and to form a power spectrum using a completely uniform weighting of all voxels, the matrix takes the form
\begin{eqnarray}
\mathbf{E}^\alpha_{ij}\Bigg{|}_{\textrm{no fg}}\!\!\!\!\!\!\!\!&=& \!\!\!\!(\mathbf{C}_{,\alpha})_{ij} \equiv \int_{V^\alpha_k} e^{i \mathbf{k} \cdot (\mathbf{r}_i - \mathbf{r}_j )} \frac{d^3 \mathbf{k}}{(2 \pi)^3} \nonumber \\
&=& \!\!\!\!\frac{1}{(2 \pi)^3} \int_0^{2\pi} \int_{k^{\perp}_{a-1}}^{k^{\perp}_{a}} \int_{k^{\parallel}_{b-1}}^{k^{\parallel}_{b}} e^{i \mathbf{k} \cdot (\mathbf{r}_i - \mathbf{r}_j )} k^{\perp} dk^{\parallel} dk^{\perp} d\varphi_k \qquad \nonumber \\
&=& \!\!\!\!\frac{(2 \pi^2)^{-1}}{(r^{\perp}_{ij})^2 r^{\parallel}_{ij}} \left( \sin k^{\parallel}_b r^{\parallel}_{ij} - \sin k^{\parallel}_{b-1} r^{\parallel}_{ij} \right) \times \nonumber \\ 
& &\left[ k_a^{\perp}  r^{\perp}_{ij} J_1(k_a^{\perp}  r^{\perp}_{ij}) - k_{a-1}^{\perp}  r^{\perp}_{ij} J_1(k_{a-1}^{\perp}  r^{\perp}_{ij})\right] \label{cylindQ},
\end{eqnarray}
where $r^{\parallel}_{ij} \equiv r_i^{\parallel} - r_j^{\parallel}$ is the radial line-of-sight distance between spatial grid points $\mathbf{r}_i$ and $\mathbf{r}_j$, $r^{\perp}_{ij} \equiv |\mathbf{r}^{\perp}| = |\mathbf{r}_i^{\perp}-\mathbf{r}_j^{\perp}| $ is the projected perpendicular distance, $\varphi_k \equiv \arccos( \widehat{\mathbf{k}}_{\perp} \cdot \widehat{\mathbf{r}}^{\perp}_{ij} ) $ is the angle between $\mathbf{k}_{\perp}$ and $\mathbf{r}^{\perp}_{ij}$, $V_k^\alpha$ is the volume in Fourier space of the $\alpha$-th bin, and $J_1$ is the $1^{st}$ Bessel function of the first kind.  (Recall that indices $a$ and $b$ specify the $k_{\perp}$ and $k_{\parallel}$ bins, respectively, and are folded into the single index $\alpha$).  The notation $\mathbf{C}_{,\alpha}$ is intended to be suggestive of the connection between the correlation function (or covariance matrix) $\mathbf{C}$ of the data and its discretized power spectrum $p_{\alpha}$:
\begin{equation}
\label{covar}
\mathbf{C} \equiv \langle (\mathbf{x}- \mathbf{m}) (\mathbf{x}-\mathbf{m})^t \rangle = \mathbf{C}_{fg} + \mathbf{N} + \sum_\alpha p_{\alpha} \mathbf{C}_{,\alpha},
\end{equation}
where in the last equality we were able to take advantage of the fact that the foregrounds, instrumental noise, and signal are uncorrelated to write the total covariance as the sum of individual covariance contributions ($\mathbf{C}_{fg}$ for the foregrounds and $\mathbf{N}$ for the instrumental noise) and the contribution from the cosmological signal (the final term) with no cross-terms.  In this form,  we see that $\mathbf{C}_{,\alpha} \equiv \partial \mathbf{C} / \partial p_{\alpha}$ is simply the derivative of the covariance with respect to the band power.  Intuitively, the last term in Equation \ref{covar} is simply an expansion of the correlation function of the cosmological function in binned Fourier modes.  As a more familiar example, consider a situation where one is trying to estimate the three-dimensional power spectrum $P_T (\mathbf{k})$ instead of $P_T (k_{\perp},k_{\parallel}) $.  In such a case, there would be no spherical or cylindrical binning in forming the power spectrum, and we would have $(\mathbf{C}_{,\alpha})_{ij} \sim e^{i \mathbf{k}_{\alpha} \cdot (\mathbf{r}_j - \mathbf{r}_i)}$.  Equation \ref{covar} then simply reduces to the well-known fact that the power spectrum is the Fourier representation of the correlation function.

Different choices for the matrix $\mathbf{E}^{\alpha}$ will give power spectrum estimates with different statistical properties.  A particularly desirable property to have is for our power spectrum estimate to be free from noise/foreground bias, so that the final estimator $\widehat{p}_\alpha$ depends only on the cosmological power and not on noise or foregrounds.  Taking the expectation value of Equation \ref{quadmethod} and substituting Equation \ref{covar} yields
\begin{equation}
\label{possiblybiasedestimator}
\widehat{p}_\alpha = \sum_\beta W_{\alpha \beta} \,p_\beta +  \textrm{tr} \left[ (\mathbf{C}_{fg} + \mathbf{N} ) \mathbf{E}^\alpha \right] - b_{\alpha},
\end{equation}
where $W_{\alpha \beta}$ is a matrix that we will discuss below.  From this expression, we see that to eliminate the noise and foreground biases, one should pick
\begin{equation}
\label{bias}
b_{\alpha} = \textrm{tr} \left[ (\mathbf{C}_{fg} + \mathbf{N} ) \mathbf{E}^\alpha \right].
\end{equation}
The presence of $\mathbf{C}_{fg}$ in this expression means that in our formalism, we can consider foreground subtraction to be a two-step process.  The first step acts on the data $\mathbf{x}$ directly through $\mathbf{E}^\alpha$ in forming the quantity $\mathbf{(x-m)}^t \mathbf{E}^{\alpha} \mathbf{(x-m)}$.  As we will see explicitly in Sections \ref{Opt} and \ref{LOS}, this step involves not just the Fourier transforming and binning as outlined above, but also a \emph{linear} foreground subtraction acting on the data.  The result is a biased first guess at the power spectrum.  The second step (where one subtracts off the $b_\alpha$ term) is a \emph{statistical} removal of foregrounds from this first guess, which acts on a \emph{quadratic} function of the data (since it is applied to the power spectrum estimate and not the data) and is similar in spirit to the methods suggested in \citep{Miguelstatistical}.  Our formalism builds on the work in \citep{Miguelstatistical} and allows one to compute the appropriate statistical foreground removal term $b_{\alpha}$ for any quadratic power spectrum estimate --- one simply plugs the corresponding $\mathbf{E}^{\alpha}$ into Equation \ref{bias}.

With $b_\alpha$ chosen appropriately, Equation \ref{possiblybiasedestimator} reduces to a relation between our power spectrum estimate and the true power spectrum:
\begin{equation}
\label{winddef}
\mathbf{\widehat{p}} = \mathbf{W} \mathbf{p},
\end{equation}
where we have grouped the components of the true power spectrum $p_\alpha$ and our power spectrum estimate $\widehat{p}_\alpha$ into the vectors $\mathbf{p}$ and $\mathbf{\widehat{p}}$ respectively\footnote{This grouping is not to be confused with our earlier grouping of data into the vector $\mathbf{x}$.  The index on $\mathbf{x}$ runs over the spatial grid points of one's measured brightness temperature distribution, whereas the index of vectors like $\mathbf{p}$ runs over different bands of $k_\perp$ and $k_\parallel$.}, and $\mathbf{W}$ is the \emph{window function matrix}, given by
\begin{equation}
\label{wind}
\mathbf{W}_{\alpha \beta} = \textrm{tr}[\mathbf{C}_{,\beta} \mathbf{E}^{\alpha} ].
\end{equation}
In general, $\mathbf{W}$ will not be a diagonal matrix, and thus each component  of our power spectrum estimate vector $\widehat{\mathbf{p}}$ (\emph{i.e.} each band power) will be a weighted sum of different components of the true power spectrum.  Put another way, the power spectrum estimate at one particular  point in the $k_\perp$-$k_\parallel$ plane is not merely a reflection of the true power spectrum at that point, but also contains contributions from the true power spectrum at nearby locations on the $k_\perp$-$k_\parallel$ plane.  Neighboring points of a power spectrum estimate are thus related to one another, and give rise to ``horizontal error bars" in one's final power spectrum estimate.  Equation \ref{wind} allows one to quantify the extent to which foreground subtraction causes unwanted mode-mixing between different parts of $k$-space, again because the $\mathbf{E}^{\alpha}$ matrix can be written to include any linear foreground subtraction process.  As a general rule of thumb, the broader one's window functions the greater the information loss in going from the true power spectrum to our power spectrum estimate.  The window functions are thus a useful diagnostic for evaluating one's estimation method, and we devote Section \ref{windfunct} to examining the window functions from various methods for estimating the power spectrum.

In addition to computing the window functions, a complete evaluation of one's power spectrum estimation technique should also involve a computation of covariance matrix $\mathbf{V}$ of the band powers:
\begin{equation}
\label{Vcovar}
\mathbf{V}_{\alpha \beta} = \langle \widehat{p}_\alpha \widehat{p}_\beta \rangle - \langle \widehat{p}_\alpha \rangle \langle \widehat{p}_\beta \rangle.
\end{equation}
Roughly speaking, the diagonal elements of $\mathbf{V}$ give the ``vertical error bars" on our power spectrum estimate.  If the signal is Gaussian, Equation \ref{Vcovar} can be written as
\begin{equation}
\label{gaussV}
\mathbf{V}_{\alpha \beta} = \sum_{ijkl} [ \mathbf{C}_{ik} \mathbf{C}_{jl}+ \mathbf{C}_{il} \mathbf{C}_{jk}] \mathbf{E}_{ij}^\alpha \mathbf{E}_{kl}^\beta.
\end{equation}

Ultimately, one of course wishes to \emph{minimize} the variances (\emph{i.e.} error bars) on our power spectrum estimate.  For a deconvolved power spectrum estimate\footnote{A power spectrum estimator that respects Equation \ref{unbiaseddef} is sometimes also said to be an unbiased estimator.  This is conceptually separate from our earlier discussion about choosing $b_\alpha$ appropriately to eliminate \emph{noise and foreground bias}.  Unfortunately, both usages of the word ``bias" are standard.  In this paper we reserve the term for the latter meaning, and instead use ``unwindowed" or ``deconvolved" when referring to the former.}, \emph{i.e.}, one where $\mathbf{W}=\mathbf{I}$ so that
\begin{equation}
\label{unbiaseddef}
\langle \mathbf{\widehat{p}} \rangle = \mathbf{p},
\end{equation}
the smallest possible error bars that can be obtained for a given experimental set-up can be computed using the Fisher matrix formalism.  The \emph{Fisher information matrix} is defined as \citep{fisher}
\begin{equation}
\label{generalfisher}
\mathbf{F}_{\alpha \beta} \equiv - \Bigg{\langle} \frac{\partial^2}{\partial p_\alpha \partial p_\beta} \ln f \Bigg{\rangle},
\end{equation}
where $f$ is the probability distribution for the data vector $\mathbf{x}$, and is dependent on both $\mathbf{x}$ and the band powers $p_\alpha$, which we defined above.  Doing so and assuming that the fluctuations are Gaussian allows one to write the Fisher matrix as
\begin{equation}
\label{powerspecfisher}
\mathbf{F}_{\alpha \beta} = \frac{1}{2} \textrm{tr}\left[ \mathbf{C}_{,\alpha} \mathbf{C}^{-1} \mathbf{C}_{,\beta}\mathbf{C}^{-1} \right].
\end{equation}
By the Cramer-Rao inequality, the quantity $(\mathbf{F}^{-1})^{1/2}_{\alpha \alpha}$ represents the smallest possible error bar on the band power $p_\alpha$  that any method satisfying Equation \ref{unbiaseddef} can achieve if one is estimating all the band powers jointly, while $(\mathbf{F}_{\alpha\alpha})^{-1/2}$ gives the best possible error if the other band powers are already known.  In our particular case, this means that if a given foreground subtraction and power spectrum estimation technique (specified by the set of $\mathbf{E}^\alpha$s) yields a covariance matrix $\mathbf{V}$ that is equal to $\mathbf{F}^{-1}$, the technique is optimal in the sense that no other \emph{unwindowed} method will be able to produce a power spectrum estimate with smaller error bars.  This method will turn out to be closely related to the inverse variance scheme that we introduce in Section \ref{Opt}, where we apply the formalism that we have developed so far to $21\,\textrm{cm}$ tomography.  Similarly, in Section \ref{LOS} we will take the traditional line-of-sight subtraction algorithms suggested by \citep{xiaomin,Judd08,paper1,paper2} and recast them in our mathematical framework.  We will see in Section \ref{results} that the traditional methods result in a substantial loss of information, resulting in error bars that are larger than one obtains with the inverse variance method.

%In practice, however, these ``best" error bars can still be prohibitively large, and better results (in the sense of smaller error bars) can be obtained by allowing $\mathbf{W} \neq \mathbf{I}$.  Having a nondiagonal $\mathbf{W}$ effectively amounts to smoothing the power spectrum in $(k_\perp,k_\parallel)$-space, which reduces the errors on the power spectrum estimate.  This is exactly what we do in Section \ref{Opt}, where we apply the formalism that we have developed so far to $21\,\textrm{cm}$ tomography and propose an algorithm where foreground subtraction is accomplished by an inverse variance weighting of the data.  We propose this as an alternative to the line-of-sight subtraction algorithms suggested by \citep{xiaomin,Judd08,paper1,paper2}, which we present in our mathematical framework in Section \ref{LOS}.  We will find in Section \ref{results} that the line-of-sight algorithms essentially ``smooth too much", and the resulting window functions are so broad that there is a substantial loss of information, resulting in error bars that are larger than one obtains with the inverse variance method.

\subsection{Inverse Variance Foreground Subtraction and Power Spectrum Estimation}
\label{Opt}
Suppose we form the quantity
\begin{equation}
\label{qbandpowers}
q_\alpha \equiv \frac{1}{2} \mathbf{(x-m)}^t \mathbf{C}^{-1} \mathbf{C}_{,\alpha}\mathbf{C}^{-1}  \mathbf{(x-m)} - b_\alpha,
\end{equation}
and take $\mathbf{\widehat{p}}  = \mathbf{F}^{-1} \mathbf{q}$ to be our power spectrum estimate.  In \citep{Maxpowerspeclossless} it was shown that this estimate is precisely the unwindowed (\emph{i.e.}, $\mathbf{W}=\mathbf{I}$), optimal estimator hinted at in the last section.  This estimator gives error bars that are exactly those specified by the Cramer-Rao bound.

In practice, however, using this estimator tends to give power spectra that are quite noisy.  
This is because window functions naturally have a width of order the inverse size of the survey volume, so insisting that  $\mathbf{W}=\mathbf{I}$ enforces a rather ill-posed deconvolution that greatly amplifies the noise.  This results in band power estimates that have error bars that are both large (despite being the ``best" allowed by Cramer-Rao) and anticorrelated between neighboring bands (see \citep{Maxgalaxysurvey2} for a more extensive discussion).  To avoid these issues, it is often preferable to smooth the power spectrum on the $k_\perp$-$k_\parallel$ plane, which mathematically means having a nondiagonal $\mathbf{W}$.  This in turn implies that we have $\mathbf{\widehat{p}} = \mathbf{W} \mathbf{p}$, which allows us to evade the Cramer-Rao bound since the $\mathbf{\widehat{p}} = \mathbf{p}$ requirement is no longer enforced.  In other words, by allowing some smoothing of our power spectrum estimate, we can construct an estimator with smaller error bars than those given by the Cramer-Rao inequality. 

In this paper, we will construct such an estimator by choosing
\begin{equation}
\label{invvar}
\mathbf{E}^{\alpha} = \frac{1}{2 \mathbf{F}_{\alpha \alpha}} \mathbf{C}^{-1} \mathbf{C}_{,\alpha}\mathbf{C}^{-1},
\end{equation}
where $\mathbf{F}_{\alpha \alpha}$ are diagonal elements of the Fisher matrix given by Equation \ref{powerspecfisher}.  Just like with Equation \ref{qbandpowers}, the choice of Equation \ref{invvar} for $\mathbf{E}^{\alpha}$ represents an inverse variance weighting of the data.  This can be seen by inserting Equation \ref{invvar} into Equation \ref{quadmethod}:
\begin{equation}
\label{normedbandpowers}
\widehat{p}_\alpha = \frac{1}{2 \mathbf{F}_{\alpha \alpha}} \mathbf{(x-m)}^t \mathbf{C}^{-1} \mathbf{C}_{,\alpha}\mathbf{C}^{-1}  \mathbf{(x-m)} - b_\alpha.
\end{equation}
Since the covariance matrix $\mathbf{C}$ is symmetric, using Equation \ref{invvar} corresponds to using an inverse variance weighted data vector ($\mathbf{C}^{-1} \mathbf{x}$) to perform our power spectrum estimate. This weighting procedure acts as our foreground subtraction step, since in Equation \ref{covar} we included the foregrounds in our covariance matrix.  The residual noise and foreground bias is subtracted by the $b_\alpha$ term, which is obtained by substituting Equation \ref{invvar} into Equation \ref{bias}.  The (nondiagonal) window functions are obtained by inserting Equation \ref{invvar} into Equation \ref{wind}, and it is by imposing the normalization condition $\mathbf{W}_{\alpha \alpha} = 1$ (as is standard) that the $(\mathbf{F}_{\alpha \alpha})^{-1}$ normalization factor appears.  Throughout this paper, Equations \ref{invvar}, \ref{normedbandpowers}, and the corresponding window functions are what we refer to as ``the inverse variance method".

In \citep{Maxpowerspeclossless}, it was shown that in the limit that the data vector $\mathbf{x}$ is drawn from a Gaussian distribution\footnote{\label{gaussianfootnote}Gaussianity is certainly \emph{not} a good assumption for $21\,\textrm{cm}$ tomography.  However, even if not strictly optimal, an inverse variance weighting of data is often desirable.  Moreover, in the presence of strongly non-Gaussian signals, the power spectrum itself is a non-optimal statistic in that it fails to capture all the cosmological information present in the raw data.  Thus, the mere act of replacing the data by a power spectrum is in this sense a Gaussian approximation, and in that limit the inverse variance method is the optimal one.  Note also that while the formulae for the error bar estimates (such as Equation \ref{invvarerrorbars}) may suffer from inaccuracies if there are non-Gaussianities, the expressions for the window functions (Equation \ref{winddef}) and the noise and foreground bias (Equation \ref{bias}) remain strictly correct.  This is because the window functions and the bias terms are derived from manipulating Equation \ref{quadmethod}, which only involves \emph{second moments} of the data vector $\mathbf{x}$ (and are therefore completely describable in terms of covariances), whereas the error bars come from Equation \ref{Vcovar}, which implicitly depends on \emph{fourth moments} of the data.}, the inverse variance method beats all other power spectrum estimators, in the sense that \emph{no other method --- windowed or not --- can deliver smaller error bars on the final power spectrum estimates}.  To compute these error bars, we insert Equation \ref{invvar} into Equation \ref{gaussV} and (assuming Gaussianity in a manner similar to \citep{Maxpowerspeclossless}), obtain
\begin{equation}
\label{optcovar}
\mathbf{V}_{\alpha \beta} = \frac{\mathbf{F}_{\alpha \beta}}{\mathbf{F}_{\alpha \alpha} \mathbf{F}_{\beta \beta}}.
\end{equation}
In particular, the ``vertical error bars" on a particular band power are given by the diagonal elements of $\mathbf{V}$, giving
\begin{equation}
\label{invvarerrorbars}
\Delta p_\alpha \equiv \mathbf{V}_{\alpha \alpha}^{1/2} = \frac{1}{\sqrt{\mathbf{F}_{\alpha \alpha}}}.
\end{equation}
This means that the inverse variance method delivers error bars that are smaller than those given by the Cramer-Rao bound, because here our error bars are equal to $(\mathbf{F}_{\alpha \alpha})^{-1/2}$, something which can only be achieved by an unwindowed estimator if all but one of the band powers are known beforehand.  The windowing has thus achieved our goal of producing a less noisy power spectrum with smaller error bars.  Note that this outcome was by no means guaranteed --- for instance, in Section \ref{results} we will find that the traditional line-of-sight methods described in Section \ref{LOS} essentially smooth the $k_\perp$-$k_\parallel$ plane too much, resulting in window functions that are so broad that there is a substantial loss of information, which in turn causes larger error bars than one obtains with the inverse variance method.

Comparing the inverse variance method to the unwindowed estimator $\mathbf{\widehat{p}} = \mathbf{F}^{-1} \mathbf{q}$ discussed earlier, we see that the difference lies in whether one normalizes the power spectrum using $\mathbf{F^{-1}}$ or $(\mathbf{F}_{\alpha \alpha})^{-1}$.  While we have just seen that the latter gives smaller error bars on the power spectrum, the choice of normalization becomes irrelevant as one goes beyond the power spectrum to constrain cosmological parameters.  This is because both choices consist of multiplying by an invertible matrix, and thus no information is lost.  This is not true for the traditional line-of-sight algorithms, where the non-optimal error bars on the power spectrum are due to an irreversible loss of information, which will in turn cause larger error bars on cosmological parameters.
\subsection{Line-of-Sight Foreground Subtraction}
\label{LOS}
In a typical\footnote{As noted above, variants of the LOS method exist, but we expect the conclusions of this paper to be independent of our specific implementation.} line-of-sight (LOS) foreground subtraction method, the measured signal from each LOS is plotted as a function of frequency (or, if one prefers, of radial distance) and a low-order polynomial is fit to the data.  The low-order fit is then subtracted from the data, with the hope that the remaining signal varies sufficiently rapidly with frequency to be dominated by the cosmological contribution to the signal and not by foregrounds (which are spectrally smooth and therefore expected to be well-approximated by low-order polynomials).  Mathematically, if one arranges the elements of the data vector $\mathbf{x}$ so that one cycles through the radial/frequency direction most rapidly and the perpendicular/angular directions less rapidly, the action of a LOS foreground subtraction can be described by the equation $\mathbf{z} = \mathbf{D} \mathbf{x}$, where $\mathbf{z}$ is the foreground-cleaned data and $\mathbf{D}$ is a block diagonal matrix.  That $\mathbf{D}$ is block diagonal is simply an expression of the fact that different lines-of-sight are independently cleaned in this algorithm, \emph{i.e.} there is no attempt to use the angular structure of foregrounds for cleaning.  If one's data cube has $n_{\parallel}$ pixels along the line of sight direction and a total of $n_{\perp}$ pixels in the perpendicular directions, then $\mathbf{D}$ consists of $n_{\perp}$ blocks, each of size $n_{\parallel} \times n_{\parallel}$ and of the form
\begin{equation}
\label{singleblock}
\mathbf{D}_{single\,block} = \mathbf{I} - \mathbf{X} [\mathbf{X}^t \mathbf{X} ]^{-1} \mathbf{X}^t,
\end{equation}
where $\mathbf{I}$ is identity matrix, $\mathbf{X}$ is an $n_\parallel \times (m +1)$ matrix such that $\mathbf{X}_{ij}$ equals the $i$th frequency (or radial pixel number) taken to the $(j-1)$th power, and $m$ is the order of the polynomial fit.  In \citep{Judd08,paper1} it was found that a \emph{quadratic} polynomial ought to be sufficient for cleaning foregrounds below the level of the expected cosmological signal.  In general, one should select a polynomial that is as low an order as possible to mitigate the possibility of accidentally removing part of the cosmological signal during foreground subtraction.

Once foreground subtraction has been performed, we can form the power spectrum by Fourier transforming, binning, and squaring.  The Fourier transform and binning are accomplished by the matrix $\mathbf{C}_{,\alpha}$, which is given by Equation \ref{cylindQ}.  Subsequently multiplying by $\mathbf{z}^t$ squares the results, so the final estimate of the power spectrum takes the form
\begin{equation}
\label{nonOPTprojectingout}
\widehat{p}_\alpha^{LOS} = \mathbf{z}^t \mathbf{C}_{,\alpha} \mathbf{z} =  \mathbf{(x-m)}^t \mathbf{D}\mathbf{C}_{,\alpha}\mathbf{D} \mathbf{(x-m)}.
\end{equation}
Comparing this to Equation \ref{quadmethod}, it is clear that the LOS foreground subtraction methods (and subsequent power spectrum estimations) proposed in the literature are quadratic methods with $\mathbf{E}^\alpha = \mathbf{D}\mathbf{C}_{,\alpha}\mathbf{D}$ and $\mathbf{b}_\alpha = 0$.  The window functions can be readily computed by substituting $\mathbf{E}^\alpha = \mathbf{D}\mathbf{C}_{,\alpha}\mathbf{D}$ into Equation \ref{wind}, giving
\begin{equation}
\mathbf{W}_{\alpha \beta}^{LOS} = \textrm{tr}[\mathbf{C}_{,\alpha} \mathbf{D}\mathbf{C}_{,\beta}\mathbf{D}].
\end{equation}
As we shall see in Section \ref{results}, the off-diagonal elements of this matrix are large, which implies that LOS foreground subtraction has the effect of introducing large correlations between the power spectrum estimates at different parts of the $k_\perp$-$k_\parallel$ plane.

Comparing the LOS algorithm to the optimal inverse variance algorithm described in Section \ref{Opt}, we can identify three areas in which the LOS algorithm is non-optimal:
\begin{enumerate}
\item \textbf{The method produces power spectra that still contain a residual noise and foreground bias.}  This conclusion has been numerically confirmed by \citep{Judd08}, \citep{LOFAR2}, and \citep{PetrovicOh}, and trivially falls out of our analytic framework --- Equation \ref{bias} shows that to eliminate the residual bias, one must subtract
\begin{eqnarray}
\label{LOSbias}
b_{\alpha}^{LOS} &=& \textrm{tr} \left[ (\mathbf{C}_{fg} + \mathbf{N} ) \mathbf{E}^\alpha \right] \nonumber \\
&=& \textrm{tr} \left[ (\mathbf{C}_{fg} + \mathbf{N} ) \mathbf{D}\mathbf{C}_{,\alpha}\mathbf{D} \right] 
\end{eqnarray}
from the initial estimate, so the setting of $\mathbf{b}_{\alpha}^{LOS}$ to zero means the method is biased.  Fortunately, since $\mathbf{D}$ and $\mathbf{C}_{,\alpha}$ are known \emph{a priori} and $\mathbf{C}_{fg}$ and $\mathbf{N}$ can be modeled (see Section \ref{foregroundsandnoise} for details), this bias can be easily removed.
\item \textbf{The method does not make full use of the available data.}  The foreground cleaning in the LOS method can be thought of as a projection of the data vector $\mathbf{x}$ into the subspace orthogonal to low order polynomials in the frequency direction\footnote{A complementary treatment of LOS foreground subtraction that also casts the subtraction as a projection of data onto a subspace of polynomials can be found in \citep{Matt3}.}.  To see that, note that the matrix $\mathbf{D}$ defined above takes the form of a symmetric ($\mathbf{D} = \mathbf{D}^t$) projection matrix ($\mathbf{D}^2 = \mathbf{D}$), so using the vector $\mathbf{z}= \mathbf{D}\mathbf{x}$ to estimate our power spectrum essentially amounts to limiting our analysis to the subset of data orthogonal to the polynomial modes.  Such a procedure, where one projects out the modes that are \emph{a priori} deemed contaminated, is exactly analogous to similar techniques in CMB data analysis (where one customarily removes the monopole and dipole modes, as well as pixels close to the Galactic plane) and galaxy survey analysis (where one might project out some purely angular modes in order to protect against incorrectly modeled dust extinction).

The projecting out of contaminated modes will necessarily result in larger error bars in one's final power spectrum estimate, because cosmological information is irreversibly destroyed in the projection procedure.  Indeed, this has been a concern with LOS subtraction, for any component of the cosmological signal that is non-orthogonal to low-order polynomials in frequency will be inadvertently subtracted from the data.  At best, if the cosmological signal happens to be completely orthogonal to the polynomial modes, estimating the power spectrum from the projected data will give error bars that are identical to an optimal method that uses the full data, since the optimal method can simply assign zero weight to the contaminated modes.

\item \textbf{The method is a non-optimal estimator even for the projected data.}  The prescription given by Equation \ref{nonOPTprojectingout} calls for a uniform weighting of the projected data $\textbf{z}$ in the estimation of the power spectrum.  In \citep{Maxgalaxysurvey1}, it was shown that ideally one should instead apply inverse variance weighting to the remaining data.  Since the covariance matrix of the reduced data $\mathbf{\tilde{C}} \equiv \mathbf{D} \mathbf{C} \mathbf{D}$ is singular\footnote{An unsurprising result, given that we have thrown away select modes in $\mathbf{x}$.}, the inverse variance weighting is accomplished using the so-called \emph{pseudoinverse}, given by
\begin{equation}
\label{pseudoinvdef}
\mathbf{M} \equiv \mathbf{D} [\mathbf{\tilde{C}} + \eta \mathbf{X}\mathbf{X}^t ]^{-1} \mathbf{D},
\end{equation}
where $\eta$ is a non-zero constant and the matrix $\mathbf{X}$ is the same as that defined for Equation  \ref{singleblock}.  This pseudoinverse can be shown to be independent of $\eta$ \citep{Maxpowerspeclossless}, and has the property that $\mathbf{\tilde{C}M}= \mathbf{I}$ in the subspace of remaining modes, as one would desire for an inverse.  To estimate a power spectrum with an inverse variance weighting of the projected data, one simply acts with the pseudoinverse after the projection:
\begin{equation}
\mathbf{z} = \mathbf{MDx} = \mathbf{Mx},
\end{equation}
where in the final step we made use of the fact that $\mathbf{D}^2 = \mathbf{D}$, so $\mathbf{MD}=\mathbf{M}$ from Equation \ref{pseudoinvdef}.

With this weighting, the resulting power spectrum estimate becomes optimal for the projected data, in that the covariance matrix $\mathbf{V}_{\alpha \beta}$ becomes equal to the expression given in Equation \ref{optcovar}, except that one uses not the full Fisher matrix $\mathbf{F}$ but the Fisher matrix of the projected data $\tilde{\mathbf{F}}$.  This Fisher matrix can be proven \citep{Maxpowerspeclossless} to take the same form as Equation \ref{powerspecfisher}, except with the pseudoinverse taking the place of the inverse covariance:
\begin{equation}
\label{reducedfisher}
\mathbf{\tilde{F}}_{\alpha \beta} = \frac{1}{2} \textrm{tr}\left[ \mathbf{C}_{,\alpha} \mathbf{M} \mathbf{C}_{,\beta}\mathbf{M} \right].
\end{equation}
\end{enumerate}
In Section \ref{results} we will use Equation \ref{reducedfisher} (inserted into Equation \ref{invvarerrorbars}) to estimate the errors for the LOS method, even though the LOS method as proposed in the literature does not optimally weight the data.  This is because once $\mathbf{M}$ has been computed, the optimal weighting can be accomplished relatively easily, and moreover most forecasts of the ability of $21\,\textrm{cm}$ tomography to constrain cosmological parameters assume that the weighting is optimal (see for instance \citep{Yi}).  The LOS method that we use to generate the results in Section \ref{results} should therefore be considered an improved version of the traditional methods found in the literature.  Even so, we will find that the inverse variance method does better.
\section{Foreground and Noise Modeling}
\label{foregroundsandnoise}
\begin{figure*}
\centering
\includegraphics[width=1.0\textwidth,trim=0cm 21.0cm 2.5cm 8.0cm, clip]{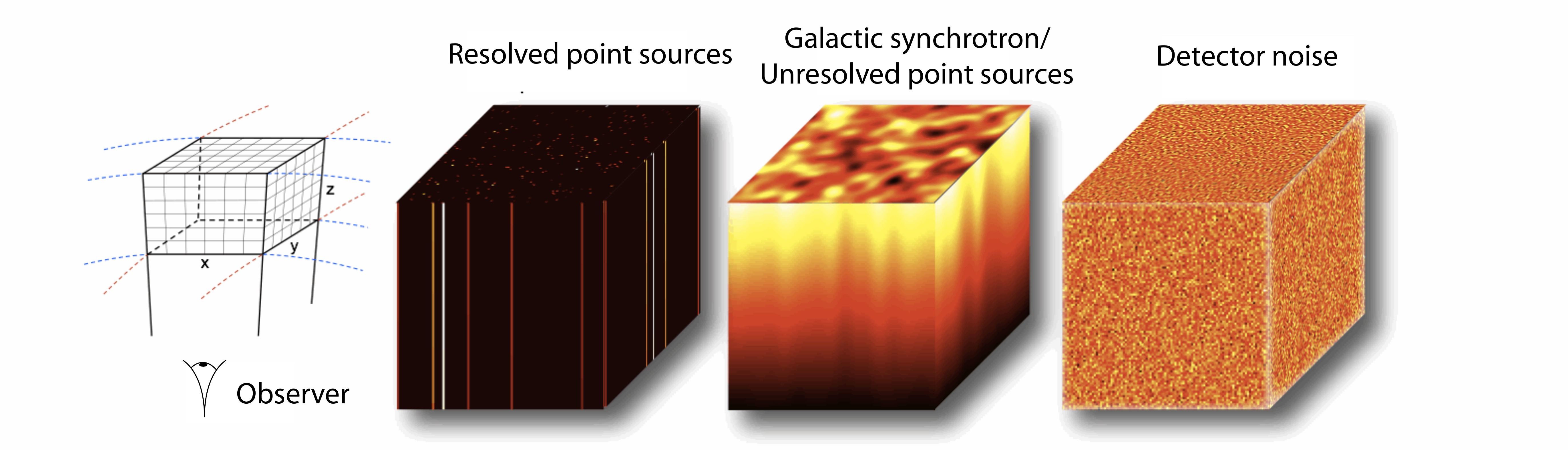}
\caption{Schematic visualization of various emission components that are measured in a $21\,\textrm{cm}$ tomography experiment.  From left to right: The geometry of a data ``cube", with the line-of-sight direction/frequency direction as the $z$-axis; resolved point sources (assumed to be removed in a prior foreground step and therefore not included in the model presented in Section \ref{foregroundsandnoise}), which are limited to a few spatial pixels but have smooth spectra; Galactic synchrotron radiation and unresolved point sources, which again have smooth spectra, but contribute to every pixel of the sky; and detector noise, which is uncorrelated in all three directions.}
\label{cubes}
\end{figure*}
Unlike the original LOS subtraction discussed in the first part of Section \ref{LOS}, the optimally weighted version of the LOS subtraction scheme and the pure inverse variance scheme of Section \ref{Opt} perform foreground subtraction in ways that are \emph{not} blind.  In other words, to inverse variance weight the data it is necessary to have a model for the covariance matrix $\mathbf{C}$, which in turn means that one must have a foreground model, since the foregrounds appear in Equation \ref{covar}.  We note, however, that since Equation \ref{winddef} depends only on geometry (via $\mathbf{C_{,\beta}}$) and one's chosen power spectrum estimation \emph{method} (via $\mathbf{E}^{\alpha}$), the window function matrix $\mathbf{W}_{\alpha \beta}$ remains strictly correct even if the foreground model is not.  In other words, even though the $\mathbf{E}^\alpha$ matrix for our inverse variance scheme involves factors of $\mathbf{C}^{-1}$, this does not detract from the accuracy of our window functions, since an incorrect $\mathbf{C}$ will simply result in a different $\mathbf{E}^\alpha$, which will in turn give different window functions that nonetheless accurately reflect what was done to the data.  It should also be noted that even with schemes where the foreground subtraction is completely blind, a proper estimation of the error bars will necessarily involve a foreground model.  We also re-emphasize that the methods presented in Section \ref{formalism} are generally applicable to \emph{any} foreground model, and that the foreground model described in this section is used only for the numerical case study presented in Section \ref{results}.

The foreground model that we use in this paper contains the following features:
\begin{enumerate}
\item Following \citep{xiaomin,Judd08,paper1,paper2}, we assume that bright, resolved point sources above some flux $S_{max}$ have been identified and removed from the data.  This may be done using traditional radio astronomy algorithms such as CLEAN \citep{CLEAN1,CLEAN2} or more recently developed techniques suitable for low-frequency radio interferometers (e.g. \citep{bernardi,Bart}).\footnote{For an analysis of how such bright point sources can affect the errors on one's final power spectrum, see \citep{dattapowerspec}.  Their analysis is complementary to what is done in this paper, in that they consider the effects of bright point source removal, whereas we consider the effects of spatially extended foregrounds.}
\item Based on extrapolations from CMB data \citep{max1}, we assume that free-free emission is negligible compared to Galactic synchrotron radiation at the relevant frequencies.  Since both free-free emission and Galactic synchrotron radiation have spectra that are well-described by the same functional form \citep{xiaomin}, we can safely ignore free-free emission for the purposes of this paper, because its contribution would be subdominant to the uncertainties inherent in the Galactic synchrotron parameters.
\item With bright point sources removed and free-free emission safely ignorable, the foreground sources that remain are unresolved point sources and Galactic synchrotron radiation.  
\end{enumerate}
Since the two foreground sources in our model are independent in origin, we may assume that their contributions to the signal are uncorrelated, thus allowing us to write the foreground covariance $\mathbf{C}_{fg}$ as the sum of an unresolved point source covariance $\mathbf{C}_{pt}$ and a Galactic synchrotron covariance $\mathbf{C}_{sync}$.  Figure \ref{cubes} provides a visualization of these various components, and in the following subsections we consider each term in turn.

\subsection{Unresolved Point Sources}
Consider first the unresolved point source covariance.  Since the unresolved point sources can still be considered ``local" in radial distance compared to the cosmological signal, the frequency dependence of these foregrounds is due entirely to the chromaticity of the sources.  This means that correlations in the LOS/frequency ``direction" are independent of the spatial distribution of the sources, and the covariance matrix can be modeled as a separable function in frequency and angular position:
\begin{eqnarray}
\label{sepcovar}
\mathbf{C}_{pt}(\mathbf{r}, \mathbf{r}^{\prime}) &\equiv&  \langle (\mathbf{x}(\mathbf{r}) -\mathbf{m}(\mathbf{r})) (\mathbf{x}(\mathbf{r}^{\prime})-\mathbf{m}(\mathbf{r}^\prime))^t \rangle \nonumber \\
&\equiv &\Gamma (\nu, \nu^\prime ) \Theta (\mathbf{r}_\perp, \mathbf{r}_\perp^\prime ).
\end{eqnarray}
The function $\Theta (\mathbf{r}_\perp, \mathbf{r}_\perp^\prime)$ encodes the spatial distribution of point sources, and is therefore where the clustering of point sources manifests itself.  While point sources are known to be clustered, the clustering is weak and for many applications it is permissible to ignore the clustering \citep{TE96}.  If clustering is ignored, the number of sources in a given pixel on the sky can be modeled using a Poisson distribution, and the resulting angular power spectrum of point sources is flat (\emph{i.e.} independent of angular scale) \citep{MaxLensing}.  This gives rise to a perpendicular correlation function/covariance $\Theta$ proportional to $\delta (\mathbf{r}_\perp - \mathbf{r}^\prime_\perp)$.

It should be emphasized, however, that it is certainly not necessary to assume that point sources are unclustered.  Whether or not clustering is included, the foreground cleaning and power spectrum estimation procedure described in Section \ref{Opt} remains the same.  One simply inserts a different covariance matrix $\mathbf{C}$ into the quadratic estimator.  In fact, the inclusion of clustering will aid foreground subtraction if the clustering pattern differs significantly from that of the cosmological signal.

Unfortunately, only future experiments will be able to probe the angular clustering of point sources on the fine arcminute scales probed by $21\,\textrm{cm}$ tomography.  For simplicity, we therefore model the $\Theta$ function as a Gaussian in the perpendicular separation $r_{\perp} \equiv | \mathbf{r}_\perp - \mathbf{r}_\perp^\prime |$.  Without loss of generality, we can normalize so that $\Theta =1$ when $\mathbf{r}_\perp = \mathbf{r}_\perp^\prime$, because the overall amplitude of the foregrounds can be absorbed into $\Gamma (\nu, \nu^\prime)$.  This means that
\begin{equation}
\label{angularpart}
 \Theta (\mathbf{r}_\perp, \mathbf{r}_\perp^\prime ) = e^{- \frac{r_\perp^2 }{2 \sigma_\perp^2}},
\end{equation}
where the $\sigma_\perp$-parameter is chosen to take the small value $\sim7\,\textrm{arcminutes}$ to reflect the weakness of the clustering.%Precise value = 7.3 armins

In contrast to the perpendicular directions, one expects correlations in the LOS direction to be strong.  In other words, $\Gamma ( \nu, \nu^\prime)$ should be large even for $\nu \neq \nu^\prime$ because of the high spectral coherence of foregrounds.  To compute this correlation, consider first a simplified model where we have a population of point sources with an average number density $n$ per steradian, giving an average of $n\Omega_{pix}$ sources per sky pixel of size $\Omega_{pix}$ steradians.  We imagine that all the point sources in this population have the same flux $S_*$ at frequency $\nu_* = 150 \,\textrm{MHz}$, and a power law frequency spectrum $\propto \nu^{-\alpha}$, with a spectral index $\alpha$ drawn from a Gaussian distribution
\begin{equation}
p ( \alpha) = \frac{1}{\sqrt{2 \pi \sigma_\alpha^2}} \exp\left[- \frac{(\alpha-\alpha_0)^2}{2 \sigma_\alpha^2}\right],
\end{equation}
with $\sigma_\alpha = 0.5$ \citep{max1} and $\alpha_0$ to be fixed later.  With this foreground model, the average signal along a particular LOS is given by
\begin{equation}
m(\nu) \equiv \langle x (\nu ) \rangle = \left(\frac{A_\nu}{\Omega_{pix}}\right) \left( n \Omega_{pix} S_*\right) \int \left( \frac{\nu}{\nu_*} \right)^{-\alpha} p (\alpha) d \alpha,
\end{equation}
where $A_\nu/\Omega_{pix}$ is a conversion factor that converts our expression from flux units to temperature units.  The $\nu$ subscript serves to remind us that it is a function of frequency, and numerically we have
\begin{equation}
\left(\frac{A_\nu}{\Omega_{pix}}\right) = 1.4\times 10^{-6} \left( \frac{\nu}{\nu_*} \right)^{-2} \left( \frac{\Omega_{pix}}{1\,\textrm{sr}} \right)^{-1}\,\textrm{mJy}^{-1}\,\textrm{K}.
\end{equation}
Similarly, the covariance (or correlation function) is
\begin{eqnarray}
\Gamma (\nu, \nu^\prime ) &\equiv& \langle (x(\nu)-m(\nu)) (x (\nu^\prime) -m(\nu^\prime))\rangle \nonumber \\
& = &\frac{A_\nu A_{\nu^\prime}}{\Omega_{pix}^2} n \Omega_{pix} S_*^2 \int   \left( \frac{\nu \nu^\prime}{\nu^2_*} \right)^{-\alpha}  p(\alpha) d\alpha. \qquad
\end{eqnarray}
Note that the covariance is proportional to $n$ and not $n^2$ because in a Poisson distribution\footnote{This is not to be confused with the spatial distribution of the sources.  What is being modeled as Poisson distributed is the number of sources in a given pixel.} the mean is equal to the variance.  Evaluating the integral gives
\begin{equation}
\Gamma (\nu, \nu^\prime) = \frac{A_\nu A_{\nu^\prime}}{\Omega_{pix}} n S_*^2 \left( \frac{\nu \nu^\prime}{\nu_*^2} \right)^{- \alpha_0 + \frac{\sigma_\alpha^2}{2} \ln \left( \frac{\nu \nu^\prime}{\nu_*^2} \right)},
\end{equation}
which shows that the superposition of power law spectra with Gaussian-distributed spectral indices gives exactly a power law with a positive running of the spectral index.

Now, in reality one of course has multiple populations of point sources with varying brightness $S_*$.  If we treat these populations as independent (in a similar fashion to what was done in \citep{MaxLensing}), then the total covariance due to all populations is given by the \emph{sum} of the covariances of the individual populations.  In the limit of an infinite number of populations, one has
\begin{equation}
\label{1Dcorrelfct}
\Gamma (\nu, \nu^\prime) = \frac{A_\nu A_{\nu^\prime}}{\Omega_{pix}} \int_0^{S_{max}} \frac{dn}{dS_*} S_*^2 dS_* \left( \frac{\nu \nu^\prime}{\nu_*^2} \right)^{- \alpha_0 + \frac{\sigma_\alpha^2}{2} \ln \left( \frac{\nu \nu^\prime}{\nu_*^2} \right)},
\end{equation}
where $dn/dS$ is the differential source count and $S_{max}$ is a maximum point source flux above which we assume the sources can be resolved and removed from the data prior to applying our foreground subtraction and power spectrum estimation scheme.  Simulations of so-called ``peeling" techniques have suggested that resolved point sources can be reliably removed down to $S_{max} \sim 10$ to $100\,\textrm{mJy}$ \citep{Bart}.  In this paper, we go a little above the middle of this range and take $S_{max} \sim 60 \,\textrm{mJy}$.  For the differential source count we use empirical fit of \citep{dimatteo1}, which takes the form
\begin{equation}
\frac{dn}{dS_*} = \left( 4.0 \, \textrm{mJy}^{-1}\,\textrm{sr}^{-1} \right) \left( \frac{S_*}{880\,\textrm{mJy}} \right)^{-1.75},
\end{equation}
and putting everything together, we obtain
\begin{equation}
\label{gammanunu}
\Gamma(\nu, \nu^\prime) = (149\,\textrm{K}^2) \left(\frac{\Omega_{pix}}{10^{-6}\,\textrm{sr}} \right)^{-1} \left( \frac{\nu \nu^\prime}{\nu_*^2} \right)^{- \alpha_{ps}+ \frac{\sigma_\alpha^2}{2} \ln \left( \frac{\nu \nu^\prime}{\nu_*^2} \right)},
\end{equation}
where $\alpha_{ps} \equiv \alpha_0 +2 = 2.5$, with this numerical value chosen\footnote{For the results presented in Section \ref{results}, we in fact used larger values for both $\alpha_{ps}$ and $\alpha_{sync}$.  We estimate the difference to be no larger than $1\%$ at any frequency.  Moreover, with larger spectral indices the foregrounds become steeper functions of frequency, and are therefore harder to subtract out, making our results more conservative.} to match extrapolations from CMB data \citep{max1}.  Combining this with Equation \ref{angularpart} gives us our expression for the unresolved point source foreground covariance matrix.

\subsection{Galactic Synchrotron Radiation}
With the synchrotron contribution, it is more difficult to write down the form of the covariance matrix, because it is unclear how to rigorously define the ensemble averages required in the computation of the covariance.  In the frequency direction, we simply use the same expression as we obtained for the unresolved point sources.  We expect this to be a reasonable model for two reasons.  Physically, the emission from unresolved point sources is also synchrotron radiation, albeit from distant galaxies.  Empirically, the spectrum of Galactic synchrotron foregrounds is well-described by the same parametric fits that work well for the unresolved point sources, except with slightly different parameters \citep{max1,xiaomin}.  We thus reuse Equation \ref{gammanunu} with $\alpha_{ps} \rightarrow \alpha_{sync}=2.8$, $\sigma_\alpha= 0.4$, and the amplitude of $\Gamma$ (\emph{including} the $\Omega_{pix}$ factor) to be $1.1\times10^5\,\textrm{K}^2$, all of which are values more suitable for Galactic synchrotron radiation \citep{xiaomin}.  As for the perpendicular part of the covariance matrix, we once again run into the same problem we faced with the unresolved point sources, namely that there is a lack of empirical spatial correlation data on the fine scales of interest.  For simplicity, we use Equation \ref{angularpart} for the Galactic synchrotron foreground contribution as well, but with $\sigma_\perp \sim 10\,\textrm{degrees}$ to reflect the fact that the Galactic synchrotron contribution is expected to be much more spatially correlated than the point sources are \citep{max1}.  %Precise value of \sigma_\perp is 6.1 degrees
\subsection{Instrumental Noise}
To model the instrumental noise contribution to the covariance matrix, we use the standard formula for the uncertainty in a measurement of the spatial wavevector $\mathbf{k}_\perp$ component of the sky brightness:
\begin{equation}
\label{noisecovar}
\widetilde{\Delta T} (\mathbf{k}_\perp, r^{\parallel} ) = \frac{\lambda^2 T_{sys}}{A_e \sqrt{\tau_k \Delta \nu}},
\end{equation}
where $\lambda$ is the observing wavelength, $\Delta \nu$ is the channel bandwidth of the instrument, $A_e$ is the effective area of an antenna, $T_{sys}$ is the system temperature of the instrument, and $\tau_k$ is the total integration time that the instrument spends observing Fourier mode $\mathbf{k}_\perp$ (see \citep{MiguelNoise} for a pedagogical discussion).  The dependence on the radial distance $r^{\parallel}$ enters via $\lambda$, whereas the dependence on $\mathbf{k}_\perp$ enters via $\tau_k$.  The amount of time $\tau_k$ that each Fourier mode is observed for is greatly affected by the layout of one's interferometer array, since each baseline in the array probes a particular Fourier mode $\mathbf{k}_\perp$ at any one instant.  Once $\widetilde{\Delta T}$ has been determined, one must take Fourier transforms in the two perpendicular directions to obtain $\Delta T$ in position space.  One can then form the quantity $\mathbf{N} \equiv \langle \Delta T ( \mathbf{r} ) \Delta T ( \mathbf{r}^\prime ) \rangle$, with the additional standard assumption that the noise is uncorrelated between different frequencies.  

In this paper, we pick fiducial system parameters that are similar to those of the MWA.  Specifically, the results in the following sections are computed for an interferometer array with $500$ antenna tiles randomly distributed with a density varying as $r^{-2}$, where $r$ is the distance from the center of the array.  The length of the maximum baseline is taken to be $1500\,\textrm{m}$, and we assume $1000\,\textrm{hrs}$ of observing time.  Rotation synthesis is performed with the telescope located at the South Pole for simplicity.  The channel width of a single frequency channel is set at $\Delta \nu = 30 \,\textrm{kHz}$, the system temperature at $T_{sys}= 440\,\textrm{K}$, and we take the effective area of each antenna tile to be $A_e = 15\,\textrm{m}^2$.  These quantities can be taken to be roughly constant over the narrow frequency range of $150$ to $150.9\,\textrm{MHz}$ (corresponding to 30 adjacent frequency channels) that we take to define the radial boundaries of the survey volume we use for the numerical case study of Section \ref{results}.  In the perpendicular directions we take our field of view to be $1.2\,\textrm{deg}^2$, which gives 16 pixels along each perpendicular direction if we take the natural pixel size of $\Omega_{pix} \approx 4.5\,\textrm{arcmin}^2$ that is suggested by our array configuration.  This results in a data cube with $16\times 16\times 30=7680$ voxels\footnote{Note that while we speak of \emph{organizing} the data into a cube (see the left-most graphic in Figure \ref{cubes}), the physical survey volume need not be cubical.  In our computations, for instance, we take into account the fact that light rays diverge, thus giving a survey volume that takes the form of a small piece of a spherical shell.  The formalism described in Section \ref{formalism} can be applied to \emph{any} survey geometry, even ones that are not contiguous.}, which is a tiny fraction of the total data output of a typical $21\,\textrm{cm}$ tomography experiment, but a reasonable amount of data to analyze for several reasons:
\begin{enumerate}
\item \textbf{Cosmological evolution of the signal.}  As we remarked in Section \ref{quadest}, the power spectrum is only a sensible statistic if translational invariance is assumed, and this in turn rests on the assumption that the cosmological evolution of the signal is negligible over the redshift range of one's data cube.  In practice this means that one must measure the power spectrum separately for many data cubes, each of which has a short radial extent.  The configuration that we use for the numerical case study in Section \ref{results} should be considered \emph{one} such small cube.
\item \textbf{Comparisons between inverse variance subtraction and line-of-sight subtraction.}  In \citep{paper1}, it was shown that line-of-sight foreground subtraction is most effective over narrow frequency ranges.  As we will discuss in Section \ref{genapp}, the inverse variance method suffers no such limitation.  However, we select a narrow frequency range in order to demonstrate (see Section \ref{results}) that the inverse variance method does a better job cleaning foregrounds even in that regime.
\item \textbf{Computational cost.}  Even with $n_{pix} =16^2 \times 30 = 7680$ voxels, the computational cost is substantial.  This is due to the fact that one must multiply and/or invert many  $7680 \times 7680$ matrices to use Equations \ref{bias}, \ref{wind}, and \ref{gaussV}.  Because of this, the results in Section \ref{results} required more than half a terabyte of disk space for matrix storage and a little over a CPU-year of computation.  The problem quickly gets worse with increasing $n_{pix}$, since a straightforward implementation of the computations involved scales as $\mathcal{O}(n_{pix}^3)$.  Fortunately, for large datasets there exist iterative algorithms for power spectrum estimation that scale as $\mathcal{O} ( n_{pix} \log n_{pix} )$ \citep{PenPowerSpec,PadPowerSpec}, and with a few approximations (such as the flat-sky approximation) these fast algorithms can be suitably adapted 
to perform unified foreground subtraction and power spectrum estimation for $21\,\textrm{cm}$ tomography as we have done in this paper \citep{josh}.
\end{enumerate}

We stress, however, that the formalism described in Section \ref{formalism} is in principle applicable to arbitrarily large survey volumes of any geometry.

\subsection{Cosmological Signal}
For the computations in Section \ref{results}, we ignore the cosmological signal.  This is justified for several reasons.  First, the signal is expected to be at least a factor of $10^4$ smaller than the foregrounds \citep{Angelica}, which means that including the signal would make little difference to the inverse variance foreground cleaning steps, as $\mathbf{C}$ would be essentially unchanged.  The line-of-sight methods, of course, remain strictly unchanged as they are blind.  Despite this, one might still object that if foreground cleaning is successful, then we should expect the cosmological signal to be the dominant contribution to the power spectrum in at least a portion of the $k_\perp$-$k_\parallel$ plane.  This is most certainly true, and the simulation of an actual measurement of a power spectrum would be incorrect without including the cosmological signal as input.  However, in this paper we do not show what a typical measured power spectrum might look like, but simply seek to quantify the errors and biases associated with our methods.  In this context, the main effect of a cosmological signal would be to add a cosmic variance contribution to our error bars at low $k_\parallel$ and $k_\perp$, but as we will see in Section \ref{results} it is precisely at the large scale modes that foreground subtraction is least successful.  The cosmic variance errors would thus be dominated by the errors induced by imperfect foreground subtraction anyway.

In short, then, it is not necessary to include the cosmological signal in our computations, and summing just $\mathbf{C}_{fg}$ and $\mathbf{N}$ yields a fine approximation to the full covariance matrix of the system $\mathbf{C}$ for our purposes.
\section{Computations and Results}
\label{results}
In this section we compute window functions, covariances, and biases for both the inverse variance and LOS methods described above.  Using Equations \ref{bias}, \ref{wind}, and \ref{gaussV}, our computations can be performed without the input of real data, and depend only on the parameters of the system.  

In what follows, we choose the $k$-bands of our power spectrum parameterization to have equal logarithmic widths.  In the $k^\parallel$ direction we have 30 bands spanning from $0.1\,\textrm{Mpc}^{-1}$ to $10\,\textrm{Mpc}^{-1}$.  In the $k^\perp$ direction we again have 30 bands, but this time from $0.01\,\textrm{Mpc}^{-1}$ to $1\,\textrm{Mpc}^{-1}$.  Note that these bands can be chosen independently of the instrumental parameters, in the sense that a quadratic estimator can be formed regardless of what $(k^\perp, k^\parallel)$-values are chosen, although the quality of the power spectrum estimate will of course depend on the instrument.
\subsection{Window Functions}
\label{windfunct}
\begin{figure}
\centering
\includegraphics[width=0.42\textwidth]{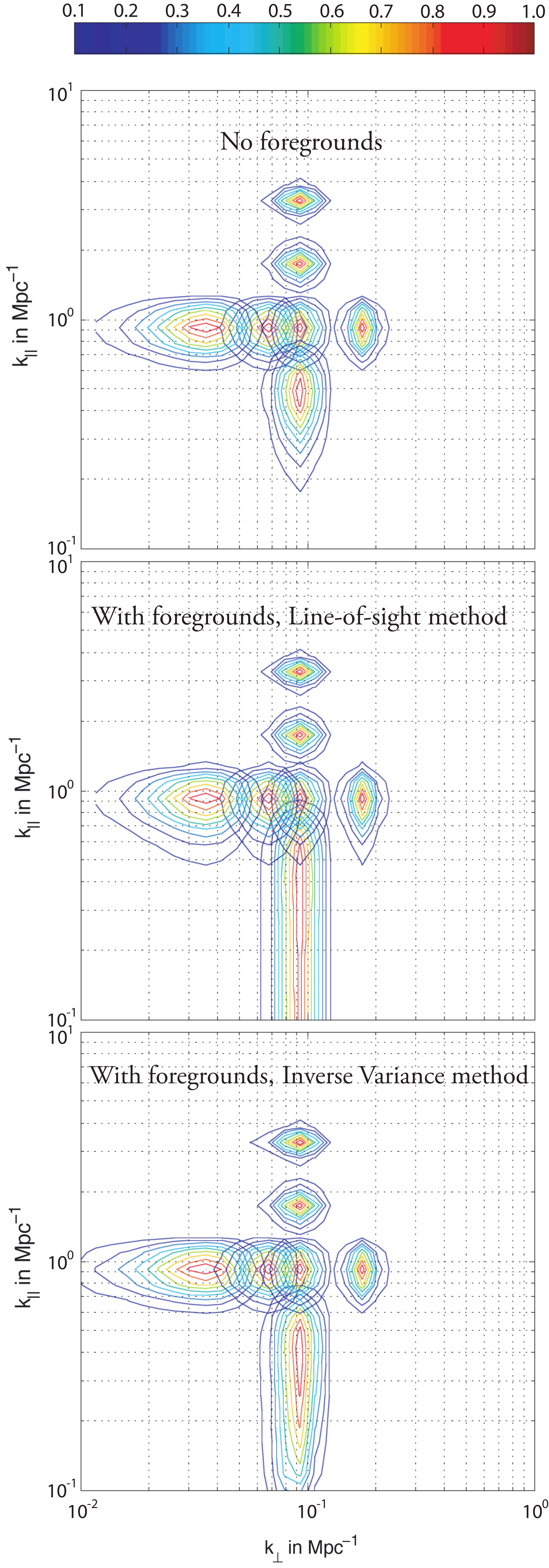}
\caption{Normalized window functions $\mathbf{\widetilde{W}}$ (see Equation \ref{windnorm}) for an unrealistic situation with no foregrounds (top panel), the line-of-sight foreground subtraction described in Section \ref{LOS} (middle panel), and the inverse variance foreground subtraction described in Section \ref{Opt} (bottom panel).}
\label{collectionofwindows}
\end{figure}
%,trim=2.0cm 7.cm 2.5cm 8.0cm, clip
%\begin{figure*}
%\centering
%\includegraphics[width=1.0\textwidth,trim=0cm 7.0cm 2.5cm 8.0cm, clip]{nonoisenofgwindcollection.pdf}
%\caption{Normalized window functions $\mathbf{\widetilde{W}}$ (see Equation \ref{windnorm}) for an unrealistic situation with no foregrounds.}
%\label{nofgnonoisewind}
%\end{figure*}
In evaluating the quality of one's power spectrum estimator, one is ultimately interested in the size of the error bars.  This means that the quantity of interest is the covariance matrix $\mathbf{V}$.  In this section, however, we first examine the window functions.  As we saw from the discussion in Section \ref{Opt}, the window functions are intimately connected to the covariances, and often provide a more detailed understanding for \emph{why} the error bars behave the way they do.  In particular, window functions that are wide and badly behaved tend to come hand in hand with large error bars.
Moreover, our expression for the window function matrix (Equation \ref{winddef}) remains strictly accurate even in the presence of non-Gaussian signals (see footnote \ref{gaussianfootnote}), so in many ways the window function matrix $\mathbf{W}$ is a more robust diagnostic tool than the covariance matrix $\mathbf{V}$.

In the top panel of Figure \ref{collectionofwindows}, we examine a situation where $\mathbf{C}=\mathbf{I}$, corresponding to a scenario where no foregrounds are present and the signal is dominated by white noise.  This is of course a highly unrealistic situation, and we include it simply to build intuition.  Rather than plotting the raw window functions, we show the normalized quantities
\begin{equation}
\label{windnorm}
\mathbf{\widetilde{W}}_{\alpha \beta} \equiv \frac{\mathbf{W}_{\alpha \beta}}{\sqrt{\mathbf{W}_{\alpha \alpha} \mathbf{W}_{\beta \beta}}},
\end{equation}
where $\mathbf{W}$ is given by Equation \ref{winddef}, and like before, each Greek index corresponds to a particular location on the $k_\perp$-$k_\parallel$ plane.  Each set of contours shown in Figure \ref{collectionofwindows} corresponds to a specific \emph{row} of $\widetilde{\mathbf{W}}$.  For example, the set of contours at the center of the ``cross" in each panel of Figure \ref{collectionofwindows} has a fixed index $\beta$ that corresponds to $(k_\perp, k_\parallel)=(0.092,0.92)\,\textrm{Mpc}^{-1}$ and is plotted as a function of $\alpha$, or equivalently, as a function of $k_\perp$ and $k_\parallel$.  From Equation \ref{winddef}, we see that each set of contours describes the weighted average of neighboring points of the true power spectrum $p$ that one is really measuring when forming the power spectrum estimate $\widehat{p}$.  With $\mathbf{C}$ set to the identity, the nonzero width of the window functions is due solely to the finite volume of our survey, and any apparent widening of the window functions towards low $k_\perp$ and $k_\parallel$ is due mostly to the logarithmic scale of our plots.
%\begin{figure*}
%\centering
%\includegraphics[width=1.0\textwidth,trim=0cm 7.0cm 2.5cm 8.0cm, clip]{LOSwindcollection.pdf}
%\caption{Normalized window functions $\mathbf{\widetilde{W}}$ (see Equation \ref{windnorm}) for the line-of-sight foreground subtraction described in Section \ref{LOS}.}
%\label{LOSwind}
%\end{figure*}

With foregrounds present, the precise shape of the window functions will depend on the foreground subtraction algorithm employed.  Shown in the middle panel of Figure \ref{collectionofwindows} are window functions for the LOS foreground subtraction described in Section \ref{LOS}.  These window functions have the same $k_\perp$-widths as those in the top panel of Figure \ref{collectionofwindows}, which is expected since the LOS foreground subtraction scheme operates only in the frequency/radial direction.  On the other hand, foreground subtraction increases the $k_\parallel$-widths of the window functions, although this effect is significant only in the lower $k_\parallel$ regions of the $k_\perp$-$k_\parallel$ plane.  At high $k_\parallel$, the window functions are very similar to those for the case with no foregrounds, since (by design) the LOS subtraction of low-order polynomials has a negligible effect on high $k_\parallel$ Fourier modes of the signal.  As one moves to lower $k_\parallel$, the foreground subtraction has more of an effect, and the window functions widen in the radial direction.  This results in large ``horizontal" error bars in the $k_\parallel$ direction for the final power spectrum estimate.  In addition, the widening of the window functions represents a loss of information at these low $k_\parallel$ values (which is unsurprising, since the very essence of the LOS method is to project out certain modes), and in Section \ref{fisherinfoanderror} we will see this manifesting itself as larger ``vertical" error bars on the power spectrum estimate.

%\begin{figure*}
%\centering
%\includegraphics[width=1.0\textwidth,trim=0cm 7.0cm 2.5cm 8.0cm, clip]{optsubtractionWindows.pdf}
%\caption{Normalized window functions $\mathbf{\widetilde{W}}$ (see Equation \ref{windnorm}) for the inverse variance foreground subtraction described in Section \ref{Opt}.}
%\label{OPTwindows}
%\end{figure*}
\begin{figure}
\centering
\includegraphics[width=0.5\textwidth,trim=3.5cm 7.5cm 3.5cm 8.0cm, clip]{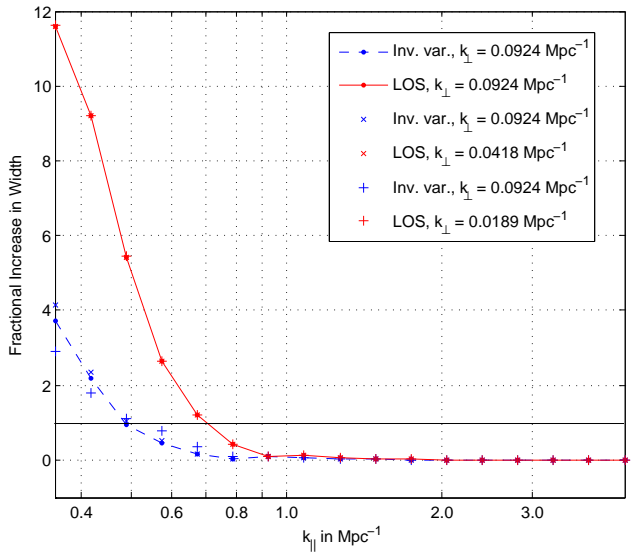}
\caption{Fractional increase (compared to the foreground-free case) in window function width in the $k_\parallel$ direction, plotted as a function of the $k_\parallel$ location (on the $k_\perp$-$k_\parallel$ plane) of the window function peak .  The blue, dashed line is for inverse variance foreground subtraction, whereas the red, solid line is for the LOS subtraction.  The various data symbols denote widths measured at different values of $k_\perp$, and the fact that they all lie close to the curves suggest that the window function width in the $k_\parallel$ direction is largely insensitive to the $k_\perp$ location of the window function peak.  The width is defined as the full-width of the window function in logarithmic $k$-space at which the function has dropped 1\% from its peak value.  The solid line is to guide the eye, and marks a fractional increase of unity \emph{i.e.} where the width is double what it would be if there were no foregrounds.}
\label{99percentwidthcompinparadirection}
\end{figure}
Shown in the bottom panel of Figure \ref{collectionofwindows} are the window functions for the inverse variance foreground subtraction described in Section \ref{Opt}.  The same elongation effect is seen in the $k_\parallel$ direction, but the problem is less severe in the sense that severe elongation does not occur until one reaches much lower values of $k_\parallel$.  This is illustrated in Figure \ref{99percentwidthcompinparadirection}, where we plot the fractional increase (compared to the foreground-free case) in window function width in the $k_\parallel$ direction for both the LOS method (red, solid line) and the inverse variance method (blue, dashed line) as a function of the $k_\parallel$ coordinate of the central peak of the window function.  The width is defined to be the full-width-99\%-max\footnote{We define the full-width-99\%-max as the full width at which the function has dropped to $99\%$ of its peak value.  A more conventional measure of the width like the full-width-half-max (FWHM) is not feasible for our purposes, as the window functions are elongated so much by foreground subtraction that if we defined the widths in such a way, the widths for the windows at the very lowest $k_\parallel$ would run off the edge of our simulation.  Our measure of the width allows a quantification of the elongation even for the lowest $k_\parallel$ values.  In all regimes where a meaningful measurement of the band powers can be made, the window functions will be compact enough for the width to be defined using the FWHM.  With the inverse variance method, for instance, one sees from the bottom panel of Figure \ref{collectionofwindows} that the FWHM remains nicely within the simulation box even though it is larger than for the foreground-free case.} (in logarithmic $k$-space) of the window in either the $k_\parallel$ or $k_\perp$ direction, and as expected there is essentially no elongation at regions of high $k_\parallel$.  Foreground subtraction begins to have an elongating effect on the $k_\parallel$ widths of the window functions at $k_\parallel < 1\,\textrm{Mpc}^{-1}$, just like we saw from our comparisons of the three panels of Figure \ref{collectionofwindows}.  The effect is much more pronounced for the LOS subtraction, and as an example we can consider the $k_\parallel$ value at which the fractional increase in width is unity (\emph{i.e.} where foregrounds cause the window functions to double in width in the $k_\parallel$ direction).  This occurs at $k_\parallel = 0.7\,\textrm{Mpc}^{-1}$ for the LOS subtraction, but not until $k_\parallel = 0.5\,\textrm{Mpc}^{-1}$ for the inverse variance subtraction, which means that with the latter scheme, one can push to lower values of $k_\parallel$ before there is significant information loss and the power spectrum band estimates become massively correlated.  Note that this conclusion holds for all $k_\perp$, as evidenced by the way the various plot symbols for different $k_\perp$ all lie very close\footnote{  For the LOS subtraction, this invariance is in fact exact since the algorithm performs the same polynomial subtraction along the line-of-sight regardless of what happens in the transverse directions.} to the curves in Figure \ref{99percentwidthcompinparadirection}.

\begin{figure}
\centering
\includegraphics[width=0.5\textwidth,trim=2.5cm 7.3cm 2.5cm 7.5cm, clip]{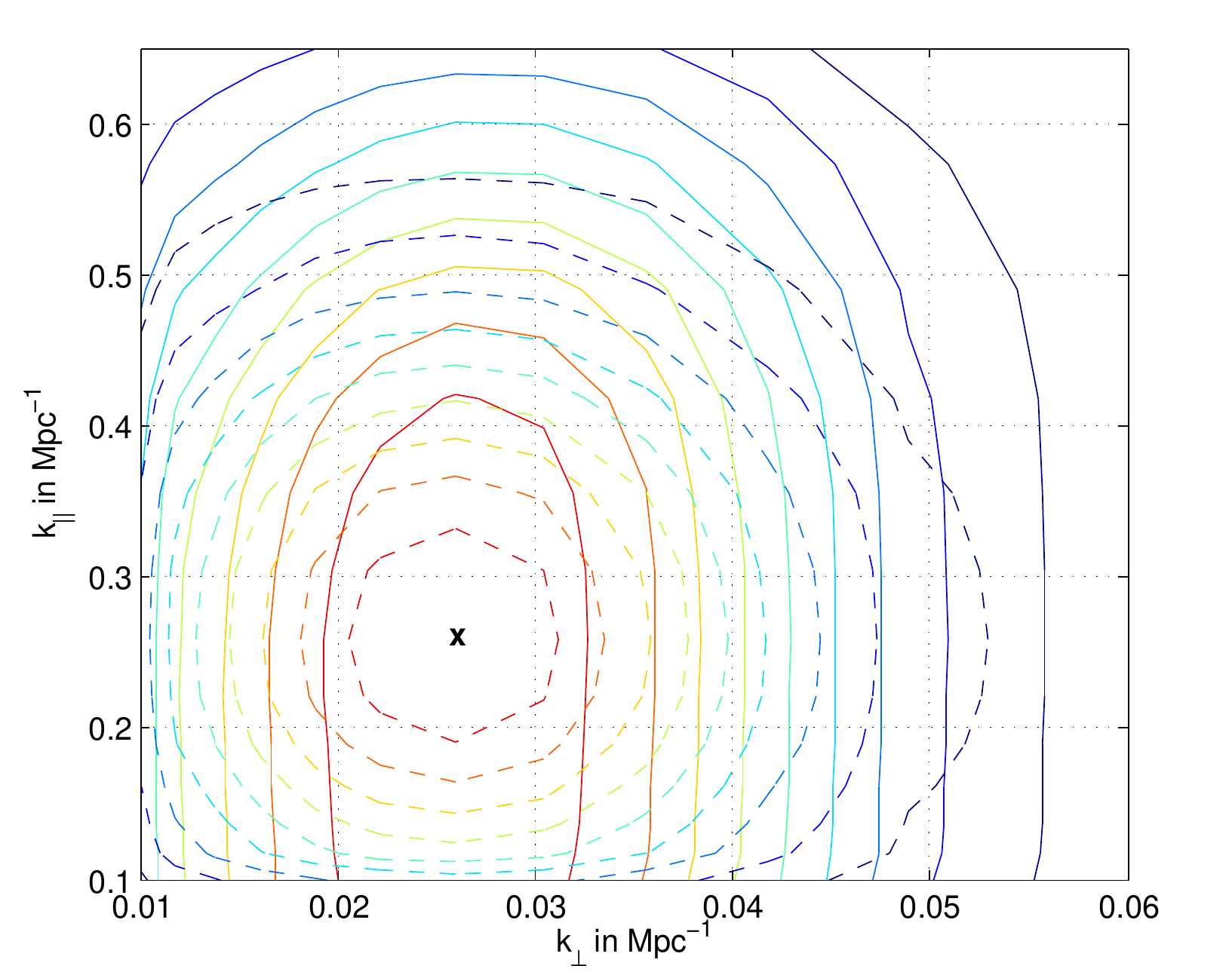}
\caption{A comparison of two normalized window functions $\mathbf{\widetilde{W}}$ centered at $(k_\perp, k_\parallel)=(0.0259,0.2593)\,\textrm{Mpc}^{-1}$ (marked by ``x").  The dotted contours correspond to a scenario with no foregrounds, while the solid contours correspond to the window functions for the inverse variance foreground subtraction described in Section \ref{Opt}.  Note the linear scale on both axes.} 
\label{lowkperpOPTcomparison}
\end{figure}
\begin{figure}
\centering
\includegraphics[width=0.50\textwidth,trim=2.5cm 7.3cm 2.5cm 7.5cm, clip]{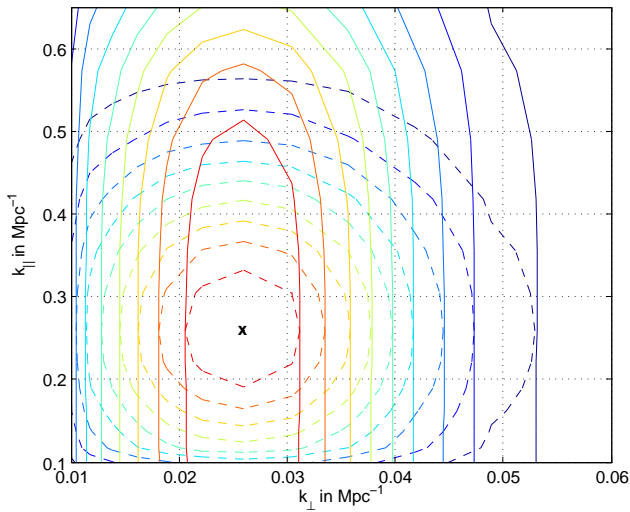}
\caption{A comparison of two normalized window functions $\mathbf{\widetilde{W}}$ centered at $(k_\perp, k_\parallel)=(0.0259,0.2593)\,\textrm{Mpc}^{-1}$ (marked by ``x").  The dotted contours correspond to a scenario with no foregrounds, while the solid contours correspond to the window functions for the LOS foreground subtraction described in Section \ref{LOS}.  Note the linear scale on both axes.} 
\label{lowkperpLOScomparison}
\end{figure}

Unlike the $k_\parallel$ widths, with the $k_\perp$ widths we find that there are important \emph{qualitative} differences in addition to quantitative differences between the two methods.  This is evident in Figure \ref{lowkperpOPTcomparison}, where we compare inverse variance and foreground-free window functions (solid and dashed contours respectively) centered at $(k_\perp, k_\parallel)=(0.0259,0.2593)\,\textrm{Mpc}^{-1}$.  The analogous plot for the LOS method is shown in Figure \ref{lowkperpLOScomparison}.  As discussed before, there is an elongation in the $k_\parallel$ direction that is more pronounced in the LOS method than the inverse variance method.  For the LOS method, this is all that happens, because the foreground subtraction is performed using only line-of-sight information.  In contrast, since the inverse variance method works by de-weighting parts of the data that are heavily contaminated by foregrounds, it is capable of also leveraging information in the transverse direction.  This results in the slight widening in the $k_\perp$ direction seen in Figure \ref{lowkperpOPTcomparison}.

\begin{figure}
\centering
\includegraphics[width=0.5\textwidth,trim=2.0cm 7.cm 2.5cm 8.0cm, clip]{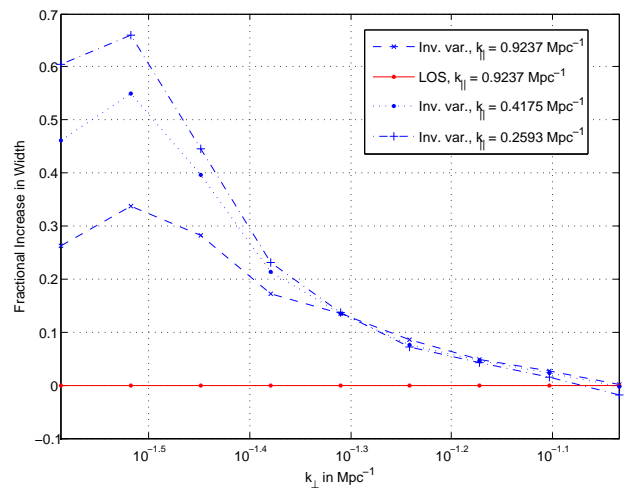}
\caption{Fractional increase (compared to the foreground-free case) in window function width in the $k_\perp$ direction, plotted as a function of the $k_\perp$ location (on the $k_\perp$-$k_\parallel$ plane) of the window function peak .  The blue, dashed line is for inverse variance foreground subtraction, whereas the red, solid line is for the LOS subtraction.  The dotted and dash-dotted curves display the same quantity as the dashed line, but at lower $k_\parallel$.  The width is defined as the full-width of the window function in logarithmic $k$-space at which the function has dropped 1\% from its peak value.}
\label{99percentwidthcompinperpdirection}
\end{figure}

In Figure \ref{99percentwidthcompinperpdirection}, we show the fractional increase in $k_\perp$ width (again as compared to the foreground-free windows) as a function of $k_\perp$.  The width increases towards lower $k_\perp$ because the inverse variance foreground subtraction is taking advantage of the smooth angular dependence of Galactic synchrotron radiation to accomplish some of the foreground cleaning.  One can also see that the width decreases with increasing $k_\parallel$, which is due again to the fact that the spectral smoothness of foregrounds mean that they dominate mainly the \emph{low} $k_\parallel$ modes, and so the inverse variance subtraction need not act so aggressively at high $k_\parallel$.

In summary, what we see is that since the inverse variance method downweights data selectively on the basis of expected foreground contamination, its approach to foreground subtraction is more ``surgical" than that of the LOS method.  For instance, at intermediate $k_\parallel$ (where foregrounds are non-negligible but certainly not at their peak), it subtracts foregrounds in a less aggressive manner than the LOS method does, allowing for final power spectrum estimates in those regions to be less correlated with each other.  This is still the case in low $k_\perp$, low $k_\parallel$ regions, but here the algorithm also uses angular information for its foreground subtraction.  Put another way, the inverse variance technique automatically ``chooses" the optimal way to subtract foregrounds.  As a result, the fact that the elongation of window functions in the $k_\parallel$ direction is so much greater than the widening in the $k_\perp$ direction (notice the different scales on the vertical axes of Figures \ref{99percentwidthcompinparadirection} and \ref{99percentwidthcompinperpdirection}) means that we now have quantitative evidence for what has thus far been a qualitative assumption in the literature --- that foreground subtraction for $21$-cm tomography is most effectively performed using line-of-sight information rather than angular information.
\subsection{Fisher Information and Error Estimates}
\label{fisherinfoanderror}

\begin{figure}
\centering
\includegraphics[width=0.5\textwidth]{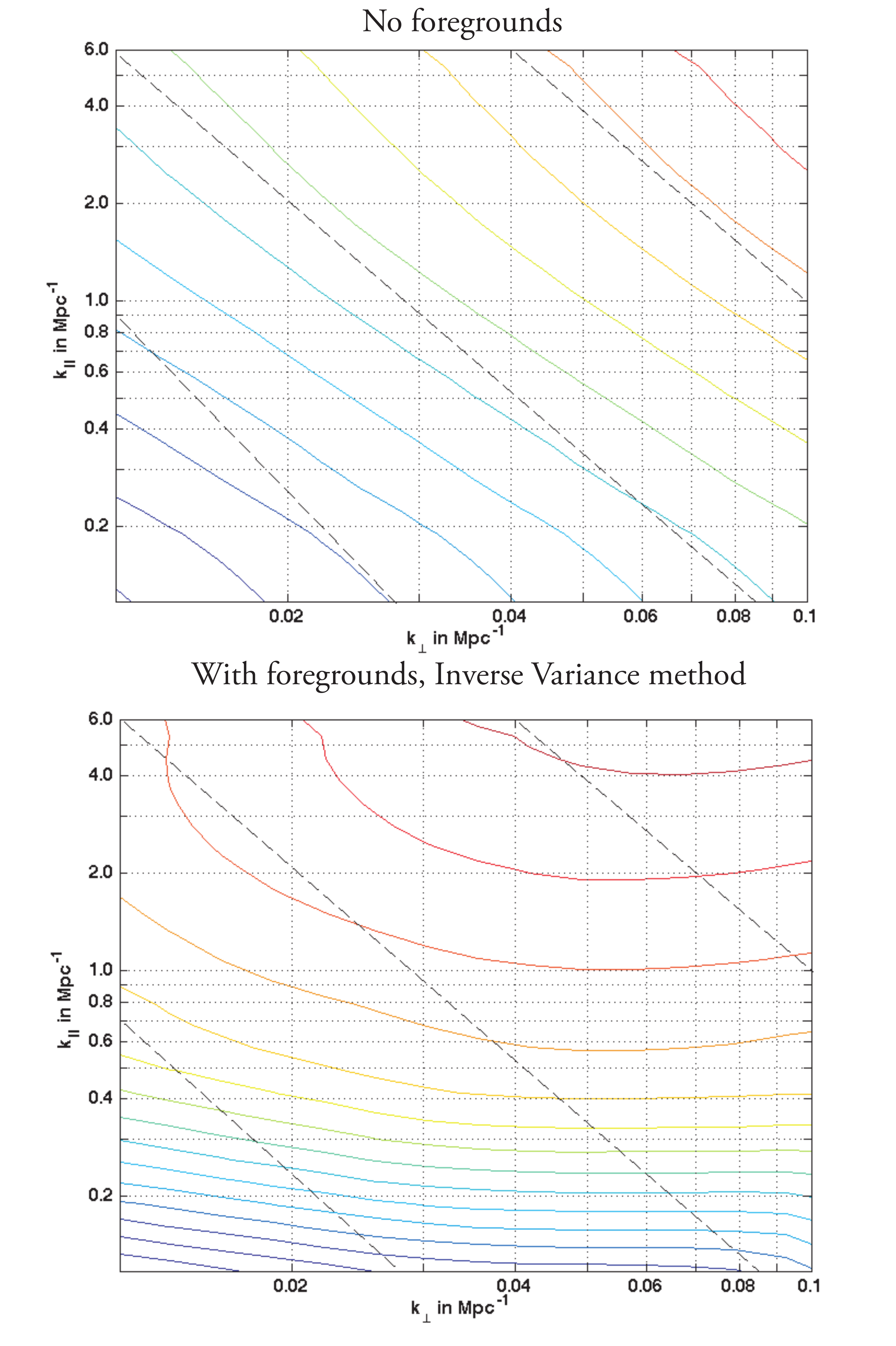}
\caption{A plot of the diagonal elements of the Fisher information matrix for an unrealistic situation with no foregrounds (top panel) and with foregrounds cleaned by the inverse variance method (bottom panel).  The (unnormalized) contours increase in value from the bottom left corner to the top right corner of each plot, and are chosen so that crossing every two contours corresponds to an increase by a factor of $10$.  Dashed lines with logarithmic slopes of $-2$ are included for reference.} 
\label{grandfisher}
\end{figure}

In the top panel of Figure \ref{grandfisher}, we return to the foregroundless scenario where $\mathbf{C}=\mathbf{I}$ and plot the diagonal elements of the Fisher matrix (which should roughly be thought of as being inversely proportional to the square of the power spectrum error bars --- see Section \ref{Opt}).  The Fisher information increases from the bottom left to the top right, and in the middle of the $k_\perp$-$k_\parallel$ plane, their contours have a (logarithmic) slope of $-2$.  To understand this, consider the process by which one forms a power spectrum in the $k_\perp$-$k_\parallel$ plane: the three-dimensional real-space data cube is Fourier transformed in all three spatial directions, and then binned and averaged over concentric annuli in $k$-space.  The averaging procedure has the effect of averaging down noise contributions to the power spectrum estimate, and thus as long as the dominant source of error is instrumental noise, the error bars on a particular part of the power spectrum will decrease with increasing annulus volume.  With $k$-bins of equal \emph{logarithmic} width (see Section \ref{results}), the volume increases twice as quickly when moving to regions of higher $k_\perp$ in the (logarithmic) $k_\perp$-$k_\parallel$ plane than when moving to regions of higher $k_\parallel$, giving a logarithmic slope of $-2$ for contours of equal Fisher information.  Conversely, Fisher information contours of slope $-2$ imply that one's error bars are dominated by instrumental noise.

At high $k_\parallel$ or high $k_\perp$, the contours shown in the top panel of Figure \ref{grandfisher} are seen to deviate from this behavior, although because of the range of the plot, the effects are somewhat visually subtle.  At high $k_\parallel$ the contours get \emph{slightly} steeper, implying a lower information content (larger error bars) than one would expect if limited by instrumental noise.  This is due to the fact that at very high $k_\parallel$ there is insufficient spectral resolution to resolve the extremely fine small scale modes, resulting in power spectrum estimates with large error bars there.  Finite angular resolution has a similar effect in the $k_\perp$ direction.

With foregrounds, the Fisher information takes a markedly different form, as seen in the bottom panel of Figure \ref{grandfisher}, where we have plotted the Fisher information after carrying out inverse variance foreground subtraction.  One sees that only small regions of the $k_\perp$-$k_\parallel$ plane have contours with a logarithmic slope of $-2$, suggesting that there are few --- if any --- parts that are dominated solely by instrumental noise.

The effects of foreground subtraction are more easily interpreted if one instead considers the following quantity\footnote{The expressions given by Equations \ref{21cmdimless} and \ref{cmbdimless} are often referred to as ``dimensionless power spectra".  This is of course somewhat of a misnomer, because both expressions carry dimensions of temperature squared, and are ``dimensionless" only in the sense that all units of length have been canceled out.}, which has units of temperature squared:
\begin{equation}
\label{21cmdimless}
\Delta_{21\,\textrm{cm}}^2 (k_\perp, k_\parallel) \equiv \frac{k_\perp^2 k_\parallel}{2 \pi^2} P_{T} (k_\perp, k_\parallel).
\end{equation}
This is exactly analogous to the quantity
\begin{equation}
\label{galaxysurveydimless}
\Delta^2 (k)\equiv \frac{k^3}{2 \pi^2} P (k)
\end{equation}
for galaxy surveys, which quantifies the contribution to the density field variance from a particular logarithmic interval in $k$, and
\begin{equation}
\label{cmbdimless}
(\delta T_\ell)^2 \equiv \frac{\ell (\ell +1)}{2 \pi} C_\ell
\end{equation}
for CMB measurements, which quantifies the contribution to temperature variance per logarithmic interval in $\ell$.  The quantity $\Delta_{21\,\textrm{cm}}^2$ thus measures the contribution to $21\,\textrm{cm}$ brightness temperature variance per logarithmic interval in $(k_\perp, k_\parallel)$-space.

\begin{figure}
\centering
\includegraphics[width=0.46\textwidth,trim=1.0cm 0.0cm 1.0cm 0.0cm, clip]{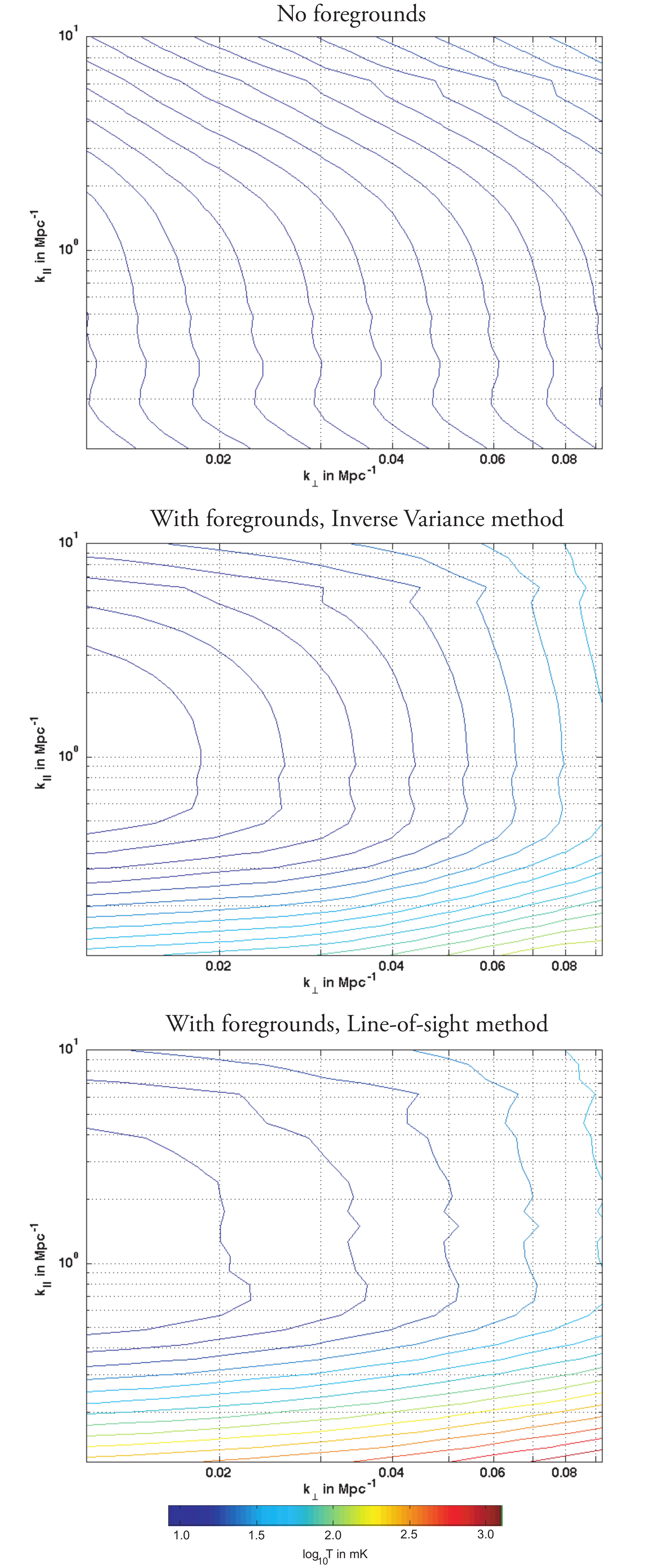}
\caption{Expected power spectrum error bars for a situation with no foregrounds (top panel, Equations \ref{powerspecfisher} and \ref{tempunitconversion} with $\mathbf{C}\propto \mathbf{I}$, scaled to match the noise-dominated $k_\perp$-$k_\parallel$ regions of the other scenarios); the inverse variance method (middle panel, Equations \ref{powerspecfisher} and \ref{tempunitconversion}); the line of sight method (bottom panel, Equations \ref{reducedfisher} and \ref{tempunitconversion}).  For the last two plots, the errors are high at low $k_\parallel$, high $k_\parallel$, and high $k_\perp$ due to residual foregrounds, limited spectral resolution, and limited angular resolution, respectively.} 
\label{tempuniterrors}
\end{figure}

Since Equation \ref{optcovar} tells us that the error bar on a particular band of $P_T(k_\perp, k_\parallel)$ is given by $(\mathbf{F}_{\alpha \alpha})^{-1/2}$ for the inverse variance method, we can estimate the error on $\Delta_{21\,\textrm{cm}}$ by computing the quantity\footnote{This expression is in fact an upper limit on the error.  That it is not exact arises from our taking of the square root of Equation \ref{21cmdimless} to obtain a quantity with temperature units.  In the limit of small errors, for instance, Taylor expanding Equation \ref{21cmdimless} with a perturbation suggests that we ought to have a factor of $1/2$ in front of our current form for $\varepsilon(k_\perp,k_\parallel)$.  It is only when the errors dominate the signal (such as at low $k_\parallel$ --- see Figure \ref{tempuniterrors}) that $\varepsilon$ tends to the expression given in Equation \ref{tempunitconversion}.  For our illustrative purposes, however, Equation \ref{tempunitconversion} suffices as a conservative estimate.}
\begin{equation}
\label{tempunitconversion}
\varepsilon (k_\perp, k_\parallel) \sim \Bigg{[} \frac{k_\perp^2 k_\parallel}{2 \pi^2} \sqrt{\frac{1}{F_{\alpha \alpha}}}\Bigg{]}^{\frac{1}{2}},
\end{equation}
which, having units of temperature, is convenient for gauging the quality of our foreground cleaning.  Note that this expression can also be used for the noiseless case (since $\mathbf{C} \propto \mathbf{I}$ can be considered a special case of the inverse variance method) as well as for the line-of-sight foreground scheme (as long as one uses the pseudoinverse and forms the Fisher matrix using Equation \ref{reducedfisher}, as outlined in Section \ref{LOS}).

Alternatively, we can think of the quantity $\varepsilon$ introduced in Equation \ref{tempunitconversion} 
as the {\it real-world degradation factor}, because it can also be interpreted as the \emph{ratio} (up to an overall dimensionful normalization factor) of the actual error on the power spectrum $P_T (k_\perp,k_\parallel)$ to the error that would be measured by a noisy but otherwise ideal experiment with infinite angular resolution and infinite spectral resolution.  In other words, $\varepsilon$ tells us how much larger the error bars get because of real-world issues like foregrounds and limited resolution.  To see this, note that such an ideal experiment would have a \emph{constant} $\varepsilon$, because the $k_\perp^2 k_\parallel$ factor essentially nulls out the geometric effects seen in the top panel of Figure \ref{grandfisher} when one is not limited by resolution.

In the top, middle, and bottom panels of Figure \ref{tempuniterrors}, we show $\varepsilon$ for the noise-only scenario, the inverse variance method, and the line-of-sight method, respectively.  With the foreground-free case in the top panel, we see that the error bars increase as one goes from the bottom-left corner (low $k_\perp$, low $k_\parallel$) to the top-right corner (high $k_\perp$, high $k_\parallel$).  The instrument's finite angular resolution causes the errors to increase in the direction of increasing $k_\perp$, whereas in the direction of increasing $k_\parallel$ this is due to the finite spectral resolution.  The smallest error bars on the plot are $\simlt 10\,\textrm{mK}$ whereas the largest are $\sim 20\,\textrm{mK}$.

Introducing foregrounds causes the errors to increase everywhere on the $k_\perp$-$k_\parallel$ plane, even after inverse variance cleaning (middle panel of Figure \ref{tempuniterrors}).  However, in addition to the effects of finite angular resolution and finite spectral resolution, we see that there are now also significant errors at low $k_\parallel$ (indeed, this is where the errors are largest).  This is due to residual foreground contamination --- being spectrally smooth, the foregrounds have their greatest effect at low $k_\parallel$, and thus that region of the $k_\perp$-$k_\parallel$ plane is the most susceptible to inaccuracies in foreground cleaning.  The ``downhill" gradients of the contours there point to higher $k_\parallel$, suggesting that the foregrounds are the dominant source of error.

Considering all the effects together, one sees that the ``sweet spot" for measurements of $\Delta_{21\,\textrm{cm}}$ (or, alternatively, where a realistic experiment compares the most favorably to an ideal experiment for measuring the power spectrum $P_T $) is at low $k_\perp$ (to avoid being limited by angular resolution) and intermediate $k_\parallel$ (to avoid being limited by spectral resolution or residual foreground contamination).  In such a region the typical errors are $\simlt 10 \,\textrm{mK}$, whereas in the resolution-limited region at the top-right corner of the plot the errors are $\sim 30 \,\textrm{mK}$ and in the foregrounds and angular resolution-limited region in the bottom-right corner the errors are $\sim 130\,\textrm{mK}$.

Although the features are seen to be qualitatively similar when we move from the inverse variance method to the LOS method (bottom panel of Figure \ref{tempuniterrors}), quantitatively we see that the errors have grown yet again, regardless of one's location on the $k_\perp$-$k_\parallel$ plane.  The smallest errors (found in the low $k_\perp$, intermediate $k_\parallel$ ``sweet spot") remain $\simlt 10\,\textrm{mK}$, and the errors in the resolution-limited region (high $k_\perp$,$k_\parallel$) remain $\sim 30\,\textrm{mK}$.  However, the foreground contaminated regions take up a larger portion of the $k_\perp$-$k_\parallel$ plane, and the most contaminated parts of the plot have errors of $\sim 900\,\textrm{mK}$.

As explained in Sections \ref{Opt} and \ref{LOS}, the higher errors are to be expected with the LOS scheme, given that the method does not make full use of the available data, having projected out the low-order polynomial modes.  Put another way, the broad window functions in the middle panel of Figure \ref{collectionofwindows} indicate that information has been lost and that power has been filtered out of the data.  Thus, in order for the power spectrum estimate to not be biased low (because of the power loss), it is necessary to multiply by a large normalization factor.  This normalization factor is derived by imposing the $\mathbf{W}_{\alpha\alpha}=1$ window function constraint (see \citep{Maxpowerspeclossless}), and manifests itself as the $\mathbf{F}_{\alpha \alpha}$ factor in the denominator of Equation \ref{normedbandpowers} --- if much power has been filtered out of the mode corresponding to the $k$-space index $\alpha$, the remaining Fisher information $\mathbf{F}_{\alpha \alpha}$ in that mode will be low, and one must divide by this small number so that the final estimate is of the right amplitude.  However, this also has the effect of magnifying the corresponding error bars, leading to the larger errors in the bottom panel of Figure \ref{tempuniterrors}.

One may be initially puzzled by the trends shown in the last two panels of Figure \ref{tempuniterrors}, which at first sight appear to differ from plots like those shown in Figure 9 of \citep{Judd08}.  In particular, in \citep{Judd08} the post-subtraction foreground residuals are the lowest at low $k_\perp$ and low $k_\parallel$, which seem to imply that the cleanest parts of the final power spectrum ought to be there.  This, however, is a misleading comparison.  A plot of residuals like that in \citep{Judd08} should be viewed as a plot of the expected size of the measured power spectrum (containing both the cosmological signal and the residual foregrounds), not the size of the \emph{error bars} associated with the measurement, which are shown in Figure \ref{tempuniterrors}.  Put another way, the power spectrum is not truly clean at low $k_\perp$ and low $k_\parallel$, for even if the measured values are small, the error bars are large, rendering that part of the power spectrum unusable.
\subsection{Residual Noise and Foreground Biases}
\begin{figure}
\centering
\includegraphics[width=0.5\textwidth]{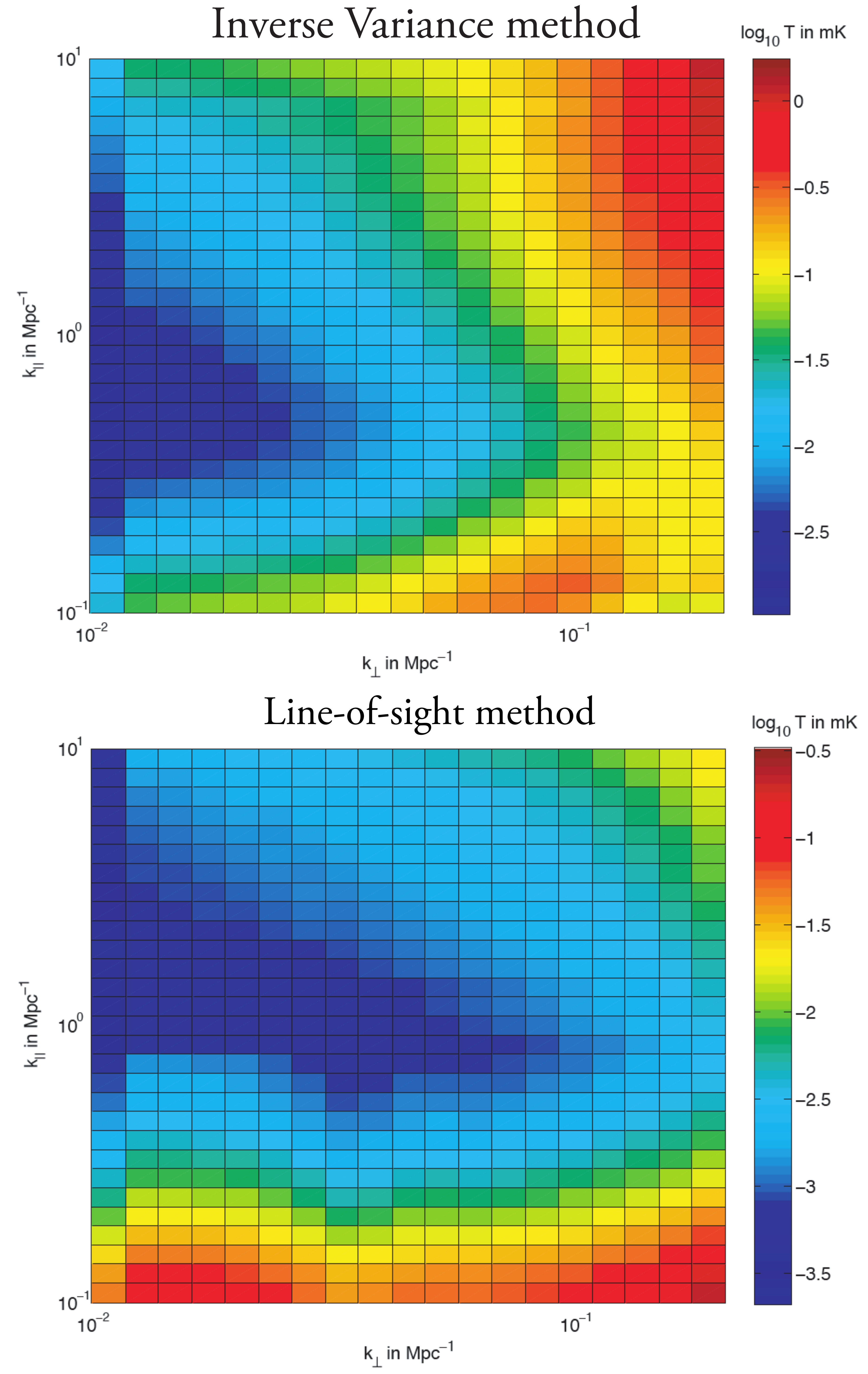}
\caption{Bias term as a function of $k_\perp$ and $k_\parallel$ for the inverse variance method (Equation \ref{bias}, top panel), and for the line-of-sight method (Equation \ref{LOSbias}, bottom panel).  The LOS plot has been artificially normalized to match the inverse variance plot at values of medium $k_\perp$ and $k_\parallel$, where we expect the two techniques to be extremely similar.}
\label{biasplots}
\end{figure}
In Figure \ref{biasplots}, we show the residual noise and foreground bias terms (Equations \ref{bias} and \ref{LOSbias}, but cast in temperature units in a way similar to Equation \ref{tempunitconversion}) for the inverse variance (top panel) and LOS (bottom panel) subtraction methods, respectively.  The biases that need to be subtracted in both methods are seen to be qualitatively similar, and the trends shown in Figure \ref{biasplots} can be readily understood by thinking of the bias subtraction in Equation \ref{quadmethod} as a foreground subtraction step in its own right.  Acting on the $k_\perp$-$k_\parallel$ plane directly, this step seeks to remove any residual foreground power from the power spectrum itself.  It therefore has its biggest effect where foreground residuals are expected to be large, and in Section \ref{toy} we will develop a simple toy model to gain intuition for how this works.

\section{An Intuitive Toy Model}
\label{toy}
In this section, we develop a simple toy model for understanding the inverse variance foreground subtraction and power spectrum estimation scheme.  The goal here is not to reproduce the exact results of the previous section, but rather, to provide intuitive explanations for how the power spectrum estimation algorithms work.

For our toy model, we consider a one-dimensional survey volume in the line-of-sight direction.  This choice is motivated by the fact that it is generally the spectral rather than spatial dependence of foregrounds that is most useful for foreground subtraction, and consequently it is the algorithm's behavior as a function of frequency that is the most interesting.  We again take the frequency range of the survey to be small.

We employ a one-dimenstional covariance of the form
\begin{equation}
C(\nu, \nu^\prime) = \delta (\nu - \nu^\prime) + N R(\nu, \nu^\prime),
\end{equation}
where we have written a continuous covariance instead of a discretized covariance matrix because we will be working in the continuous limit in this section in order to eliminate pixelization effects.  The foregrounds are modeled by $R(\nu, \nu^\prime)$, $N$ is a normalization constant to be determined later, and the noise is modeled by the delta function.  Without loss of generality, we can select units such that the r.m.s. foreground covariance is equal to unity at all frequencies \emph{i.e.} $R(\nu,\nu)=1$ for all $\nu$.  This amounts to dividing the line-of-sight covariance $\Gamma (\nu, \nu^\prime)$ of Section \ref{foregroundsandnoise} by a frequency-dependent normalization $\sigma(\nu) \sigma(\nu^\prime)$, where
\begin{equation}
\sigma (\nu) = \left( \frac{\nu}{\nu_*} \right)^{- \alpha_0 + \sigma_\alpha^2 \ln \left( \nu / \nu_*\right) },
\end{equation}
giving
\begin{equation}
R (\nu, \nu^\prime) = \exp \left[ - \frac{\sigma_\alpha^2}{2} \left( \ln \nu - \ln \nu^\prime \right)^2\right] ,
\end{equation}
\emph{i.e.} a Gaussian in logarithmic frequency.  Since we are assuming a narrow frequency range, the logarithmic terms can be expanded about $\nu_*$, giving
\begin{equation}
 R(\nu, \nu^\prime) \approx R (\nu- \nu^\prime) = \exp \left[ -  \frac{(\nu -\nu^\prime)^2}{2 \nu_c^2}\right],
\end{equation}
where we have defined the \emph{foreground coherence length}\footnote{While this definition of the foreground coherence length is what follows from Taylor expanding $R(\nu,\nu^\prime)$, we caution that Equation \ref{badcoherenceform} should not be taken too literally as a way to compute $\nu_c$.  Doing so gives rather long coherence lengths, over which the linearized version of $R(\nu,\nu^\prime)$ becomes a bad approximation.  Indeed, it is somewhat unphysical for $\nu_c$ to depend on $\nu_*$, which after all was just an arbitrary frequency in the power-law foreground parameterizations of Section \ref{foregroundsandnoise}.  In what follows, we will choose a fiducial value of $\nu_c = 0.5\,\textrm{MHz}$ to ``anchor" our toy model to the results of Section \ref{results}.}:
\begin{equation}
\label{badcoherenceform}
\nu_c \equiv \frac{1}{\sigma_\alpha}\left(\frac{\nu_*}{ \ln \nu_*}\right).
\end{equation}
Note that the foreground contribution now depends on just the \emph{difference} between $\nu$ and $\nu^\prime$ \emph{i.e.} it is translation invariant in frequency space.  The same is true for the delta function noise term, which accurately captures the fact that the instrumental noise is uncorrelated between different frequencies.  For simplicity, we model the noise as being white in our chosen units\footnote{This is of course not strictly correct, but suffices for the illustrative purposes of this toy model, especially given the narrow frequency ranges required for power spectrum estimation in $21\,\textrm{cm}$ tomography.}.  Putting everything together, we see that our toy model is one that is defined by the covariance
\begin{equation}
\label{finaltoycovar}
C ( \nu - \nu^\prime ) = \delta (\nu - \nu^\prime) + \frac{\gamma }{\sqrt{2 \pi \nu_c^2}} \exp \left[ -  \frac{(\nu -\nu^\prime)^2}{2 \nu_c^2}\right],
\end{equation}
where we have redefined our normalization constant so that the noise and foreground pieces are individually normalized to unity, allowing the constant $\gamma$ to be interpreted as a foreground-to-noise ratio.

\begin{figure}
\centering
\includegraphics[width=0.48\textwidth]{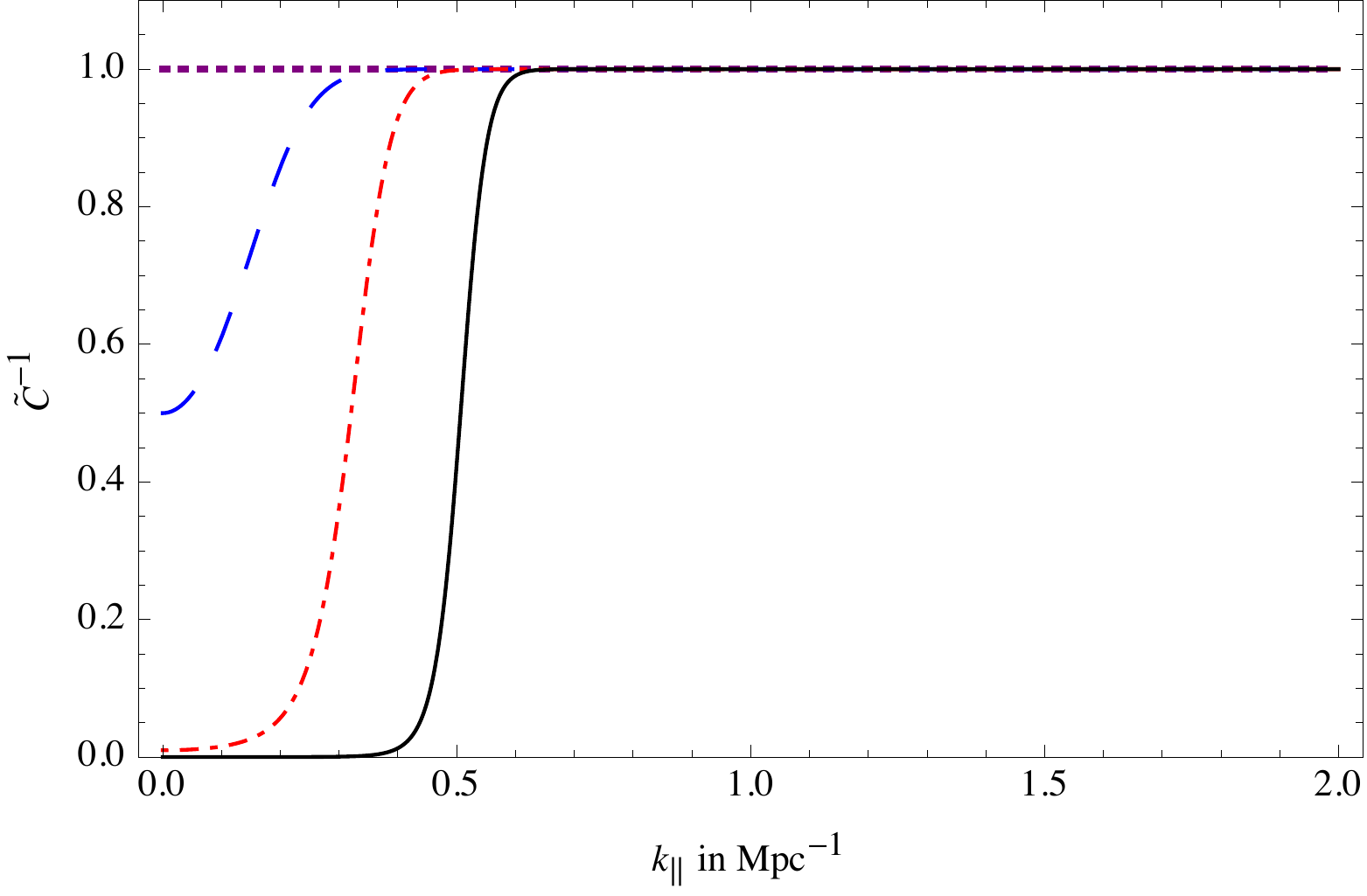}
\caption{A plot of the foreground cleaning kernel $\tilde{C}^{-1}$ (Equation \ref{invkernel}) for different values of the foreground-to-noise ratio $\gamma$, set at $\gamma=0$ for the purple/dotted curve, at $\gamma=1$ for the blue/dashed curve, at $\gamma = 100$ for the red/dot-dashed curve, and at $\gamma = 10^5$ for the black/solid curve.  In all cases, the foreground coherence length $\nu_c=0.5\,\textrm{MHz}$.  The black/solid curve is intended to be representative of a first-generation $21\,\textrm{cm}$ tomography experiment.} 
\label{FDgamma}
\end{figure}

\begin{figure}
\centering
\includegraphics[width=0.48\textwidth]{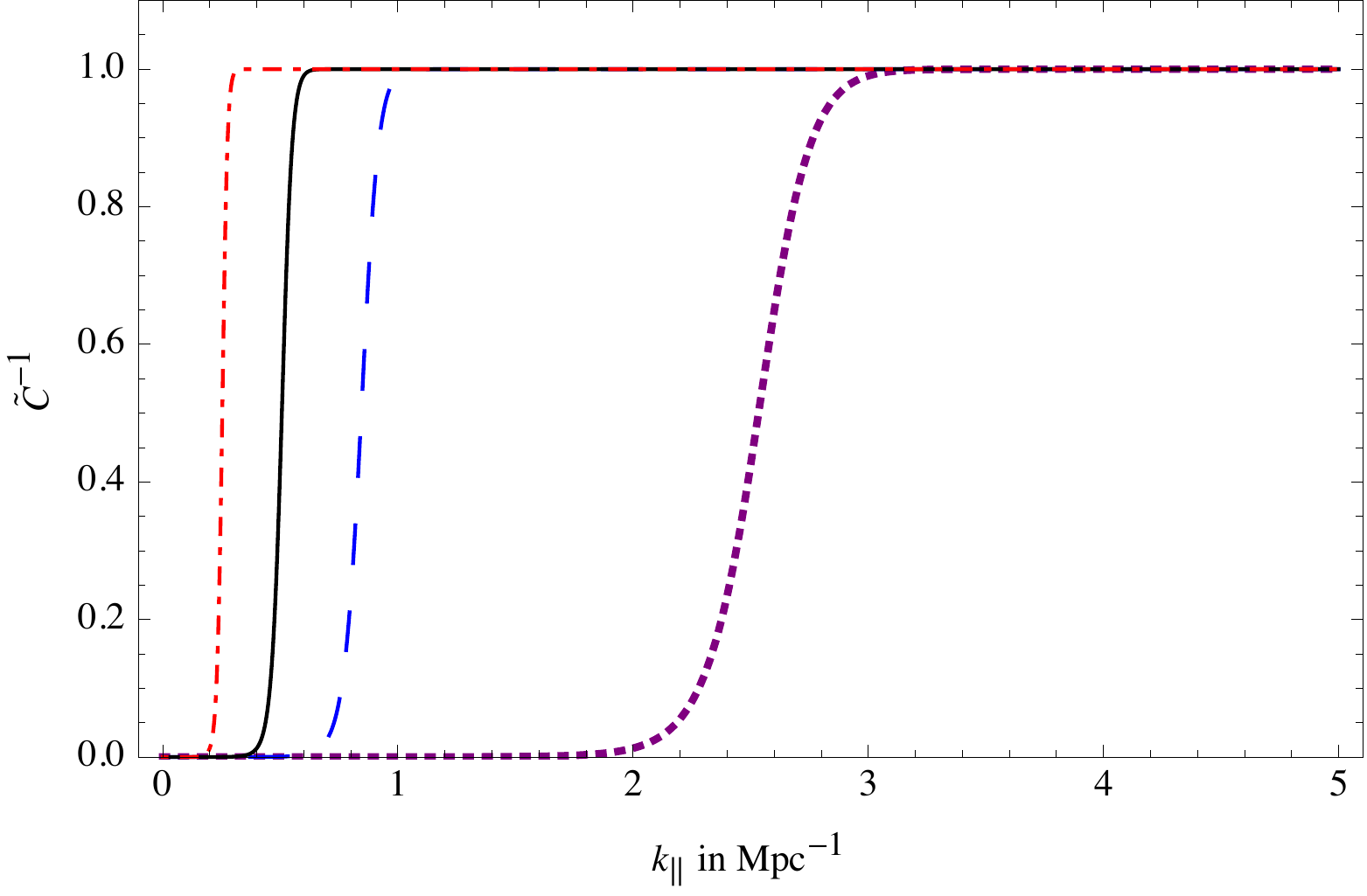}
\caption{A plot of the foreground cleaning kernel $\tilde{C}^{-1}$ (Equation \ref{invkernel}) for different values of the foreground coherence length $\nu_c$, with $\nu_c=0.1 \,\textrm{MHz}$ for the purple/dotted curve, at $\nu_c=0.3 \,\textrm{MHz}$ for the blue/dashed curve, at $\nu_c=0.5 \,\textrm{MHz}$ for the black/solid curve, and at $\nu_c=1.0 \,\textrm{MHz}$ for the red/dot-dashed curve.  In all cases the foreground-to-noise ratio $\gamma$ is fixed at $10^5$.  The black/solid curve is intended to be representative of a first-generation $21\,\textrm{cm}$ tomography experiment.} 
\label{FDnuc}
\end{figure}

Let us now apply the inverse-variance power spectrum estimation method of Section \ref{Opt} to our toy model.  Interpreting $C(\nu- \nu^\prime)$ as an integral kernel, an inverse-variance weighting of the data corresponds to applying the inverse kernel $C^{-1} (\nu - \nu^\prime)$.  By the convolution theorem, a translation invariant kernel acts multiplicatively when written in its dual Fourier basis\footnote{Strictly speaking, this is only true if the kernels are integrated from $-\infty$ to $\infty$, which is somewhat unphysical since $\nu >0$ and moreover must be within the frequency range of one's instrument.  However, since the foreground kernel decays quickly away from $\nu=\nu^\prime$, we make this approximation for the purposes of our toy model}.  This makes the computation of the inverse kernel straightforward --- one simply computes the reciprocal of the Fourier transform of $C$.  Defining $\eta$ to be the Fourier dual of $\nu$, the inverse kernel can be shown to take the form of a logistic curve in $\eta^2$:
\begin{eqnarray}
\label{invkernel}
\tilde{C}^{-1} &=& \left[1+\gamma \exp \left( - \frac{1}{2} \nu_c^2 \eta^2 \right) \right]^{-1} \nonumber \\
&\approx & \left[1+\gamma \exp \left( - \frac{\nu_c^2 c^2  k_\parallel^2}{2 \nu_* \nu_0 H_0^2 \Omega_m}   \right) \right]^{-1}, \label{invkernel2}
\end{eqnarray}
where $\nu_0 = 1420\,\textrm{MHz}$ is the rest frequency of the $21\,\textrm{cm}$ line, and $\Omega_m$, $H_0$, and $c$ have their usual meanings.  This kernel is plotted in Figures \ref{FDgamma} and \ref{FDnuc}. In Figure \ref{FDgamma}, we keep the foreground coherence length $\nu_c$ fixed at $0.5 \,\textrm{MHz}$ (chosen to calibrate our toy model against the results of Section \ref{results}) and vary the foreground-to-noise ratio $\gamma$ from 0 to $10^5$, whereas in Figure \ref{FDnuc} we keep $\gamma$ fixed at $10^5$ (typical of first-generation $21\,\textrm{cm}$ tomography experiments with $1000\,\textrm{hrs}$ of integration time) and vary $\nu_c$.  In both instances the fiducial experimental cases are plotted using solid black curves.

From the plots, ones sees that the action of the foreground cleaning kernel $\tilde{C}^{-1}$ is to suppress low-$k_\parallel$ modes in the final power spectrum estimate.  Foreground subtraction in the inverse variance scheme thus amounts to applying a high-pass filter in the frequency/line-of-sight direction.  The highest $k_\parallel$ modes are left untouched, whereas the lowest $k_\parallel$ modes are suppressed by a factor of $1+\gamma$.  This can be seen in Figure \ref{FDgamma}, where there is no suppression when $\gamma = 0$.  Physically, this is because with $\gamma=0$ there are no foregrounds and one only needs to deal with instrumental noise.  Since noise contributions from different frequencies are uncorrelated, the effect of noise is best mitigated through simple averaging and binning, which is accomplished by the $\mathbf{C}_{,\alpha}$ piece of Equation \ref{normedbandpowers} and not by the inverse variance weighting.  The $\tilde{C}^{-1}$ kernel is thus flat when $\gamma=0$.  At high values of $\gamma$, the low-$k_\parallel$ modes are essentially taken out completely because those are precisely the modes that are heavily contaminated by foregrounds.  In intermediate regimes, there is partial suppression of the low-$k_\parallel$ modes to mitigate the foregrounds, but a complete suppression is unnecessary because the foregrounds are relatively low in amplitude; thus, useful cosmological information may still be extracted from these modes.

To understand the dependence of $\tilde{C}^{-1}$ on the coherence length $\nu_c$, it is useful to define a cutoff scale $k_\parallel^{cut}$.  We define this scale to be the scale at which the $k_\parallel$ modes are ``half-suppressed", with a weighting that is the arithmetic mean of $(1+\gamma)^{-1}$ (the weighting at $k_\parallel=0$) and 1 (the weighting at $k_\parallel= \infty$).  A little bit of algebra reveals that
\begin{equation}
\label{etacut}
k_\parallel^{cut} = \left(\frac{c}{H_0}\right)^{-1} \left(\frac{\nu_* \nu_0}{\nu_c^2}\right)^{\frac{1}{2}} \Omega_m^{1/2} \sqrt{2 \ln (2 + \gamma) }.
\end{equation}
For $k_\parallel \ll k_\parallel^{cut}$, the modes are severely contaminated by foregrounds and are largely thrown out, whereas for $k_\parallel \gg k_\parallel^{cut}$ the foregrounds are minimal and essentially no cleaning is required.  Equation \ref{etacut} and Figure \ref{FDgamma} show that as foregrounds become increasingly important (\emph{i.e.} as $\gamma$ grows), $k_\parallel^{cut}$ increases.  This is because with higher foregrounds, one is forced to clean to higher $k_\parallel$ modes before estimating the power spectrum.  Encouragingly, however, $k_\parallel^{cut}$ depends only \emph{logarithmically} on $\gamma$, which means that as the foregrounds become increasingly important, only a few more modes need to be thrown out.  For instance, as the foreground-to-noise ratio $\gamma$ goes from $1$ to $10^5$, $k_\parallel^{cut}$ increases by only a factor of $\sim 3$.  In Figure \ref{FDnuc} and Equation \ref{etacut} we see that as $\nu_c$ increases, $k_\parallel^{cut}$ decreases.  Physically, this is because a high $\nu_c$ implies foregrounds that are very spectrally coherent, which allows the foreground cleaning to be confined to a handful of large scale (small $k_\parallel$) modes.

As noted in Section \ref{formalism}, subtracting foregrounds using the inverse variance method is a two-step process.  The first step is a linear one that acts on the data itself, and we have seen in this section that this first step can be roughly understood as a high-pass filter in the frequency direction.  The second step involves foreground mitigation in the power spectrum itself and amounts to the subtraction of the residual noise and foreground bias term $b_\alpha = \textrm{tr}[\mathbf{C}_{,\alpha} \mathbf{C}^{-1}]$.  This term can be approximated in our toy model by replacing the discrete trace by a continuous integral:
\begin{eqnarray}
b_\alpha &=& \textrm{tr}[\mathbf{C}_{,\alpha} \mathbf{C}^{-1}] = \sum_{i,j} (\mathbf{C}_{,\alpha})_{ij} \mathbf{C}^{-1}_{ji}  \label{optbias2} \nonumber \\
&\approx &\int C_{,\alpha} (\nu_i,\nu_j) C^{-1} (\nu_j, \nu_i) d\nu_i d\nu_j \nonumber \\
&\propto & \int e^{i \eta_\alpha (\nu_i - \nu_j) } C^{-1} (\nu_j, \nu_i) d\nu_i d\nu_j = \tilde{C}^{-1} (\eta_\alpha),\qquad
\end{eqnarray}
where we used the fact that in one dimension, $C_{,\alpha} (\nu_i,\nu_j) \propto e^{i \eta_\alpha (\nu_i - \nu_j) } $.  With the final integral being precisely the Fourier transform of $C^{-1}$, we see that the bias term that needs to be subtracted off is proportional to the $\tilde{C}^{-1}$ function given by Equation \ref{invkernel}.  Intuitively, this can be thought of as a subtraction of the statistically expected foreground residuals in $k_\parallel$-space.  From the form of Equation \ref{invkernel2}, we can see that at low $k_\parallel$ --- where the linear foreground subtraction step acts more aggressively --- the residuals are expected to be small, and the extra foregrounds that need to be removed in $k_\parallel$-space are small.

Using similar manipulations, one can see that the bias term acts in a similar way for the LOS polynomial subtraction.  From Equation \ref{LOSbias} we have $b_\alpha = \textrm{tr}[\mathbf{C}_{,\alpha} \mathbf{D} \mathbf{C} \mathbf{D} ]$ for the LOS subtraction, where recall that $\mathbf{D}$ was the projection matrix responsible for projecting out the lowest order polynomials in the frequency direction (which hopefully contain most of the foregrounds).  Comparing this to Equation \ref{optbias2} tells us that one simply needs to substitute $\mathbf{C}^{-1}$ for $\mathbf{D C D}$.  This combination can be rewritten as
\begin{equation}
\mathbf{DCD} = \langle \mathbf{D} (\mathbf{x}- \mathbf{m}) (\mathbf{x}- \mathbf{m})^t \mathbf{D}^{t} \rangle,
\end{equation}
which is the covariance of the residual foregrounds after LOS subtraction.  Conceptually, then, the residual bias term works in the same way as it did in the inverse variance subtraction --- one is simply subtracting off the statistically expected foreground residuals in power spectrum space.

Finally, we can make a similar estimate of the Fisher information for the inverse variance method.  Starting from Equation \ref{powerspecfisher}, we have
\begin{eqnarray}
\mathbf{F}_{\alpha \beta} &=& \frac{1}{2} \textbf{tr}\left[ \mathbf{C}_{,\alpha} \mathbf{C}^{-1} \mathbf{C}_{,\beta} \mathbf{C}^{-1} \right] \nonumber \\
&\approx & \frac{1}{2} \int C_{,\alpha} (\nu_i,\nu_j) C^{-1} (\nu_j, \nu_k) \times \nonumber \\
 & &\quad C_{,\beta} (\nu_k,\nu_m) C^{-1} (\nu_m, \nu_i)  d\nu_i d\nu_j d\nu_k d\nu_m\quad \nonumber \\
 &\propto & \Bigg{|} \int e^{-i \eta_\alpha \nu_j} C^{-1} (\nu_j, \nu_k) e^{i \eta_\beta \nu_k} d\nu_j d\nu_k \Bigg{|}^2 \nonumber \\
& \propto &\Big{|} \tilde{C}^{-1} ( \eta_\alpha ) \Big{|}^2 \propto \frac{\delta_{\alpha \beta}}{\left[ 1+ \gamma \exp\left( - \frac{1}{2} \nu_c^2 \eta_\alpha \right)\right]^2} \nonumber \\
&\approx & \delta_{\alpha \beta} \left[1+\gamma \exp \left( - \frac{\nu_c^2 c^2  k_\parallel^2}{2 \nu_* \nu_0 H_0^2 \Omega_m}   \right) \right]^{-2}, \label{toyfisher}
\end{eqnarray}
which looks exactly like the curves plotted in Figure \ref{FDgamma}, but squared.  This shows that the Fisher information content is much higher at high values of $k_\parallel$ than at low values of $k_\parallel$.  In other words, after foreground subtraction removes power from large scales (small $k_\parallel$) there is little information left there and the error bars are large.  This is in accordance with the middle panel of Figure \ref{tempuniterrors}, where the expected errors are seen to decrease as one moves away from the lowest $k_{\parallel}$ values.  Note, however, that the finite spectral resolution effects that we saw in the high $k_\parallel$ regions of Figure \ref{tempuniterrors} are \emph{not} captured by Equation \ref{toyfisher}.  This is to be expected, since here we have a toy model whose covariance is a \emph{continuous} function of frequency.

The results of this section suggest that the inverse variance foreground subtraction can be qualitatively thought of as a scheme where the data are high-pass filtered in the frequency direction.  This can be done  by applying a multiplicative filter in $k_\parallel$ space, an example of which is given by Equation \ref{invkernel2}.  In fact, such line-of-sight filtering can be considered a computationally cheap approximation to the full inverse variance method, requiring just a simple multiplication instead of the inversion of an $n_{pix} \times n_{pix}$ covariance matrix.  The results should be almost as good, since we saw from the bottom panel of Figure \ref{collectionofwindows} that even in the full method, the algorithm ``chooses" to perform foreground subtraction almost exclusively in the line-of-sight direction.

Filtering in the line-of-sight direction is an approach that is extremely similar to one that was proposed in \citep{PetrovicOh}.  There, the authors suggest cleaning foregrounds by excluding all modes with $k_\parallel < k_{\parallel}^{crit}$, where $k_{\parallel}^{crit}$ is some critical line-of-sight mode below which foregrounds are expected to dominate.  This corresponds to applying a step-function filter, and in \citep{PetrovicOh} it was found that this dramatically reduces systematic biases in the spherically averaged 3D power spectrum.  However, the authors also caution that some foregrounds may have some small (but nonzero) high $k_\parallel$ component that will not be alleviated by a sharp filter.  This problem is solved by instead using the filter $\tilde{C}^{-1} (k_\parallel) $ given in Equation \ref{invkernel2}.  From Figures \ref{FDgamma} and \ref{FDnuc} one sees that the filter is qualitatively similar to a step function, but the fact that the filter is itself dictated by a foreground model allows it to enact slight suppressions of foregrounds even at high $k_\parallel$ modes.

\section{Discussion and Conclusions}
\label{conc}
In this paper, we have presented a unified matrix-based formalism for $21\,\textrm{cm}$ foreground removal and power spectrum estimation.  Using this framework, we compared existing line-of-sight (LOS) polynomial foreground subtraction schemes with a proposed inverse variance method, quantifying their errors and showing how to eliminate their biases.  We now review our basic results, and follow with a discussion of their general applicability before summarizing our conclusions.
\subsection{Basic Results}
Through the numerical case study performed in Section \ref{results}, we have established a number of qualitative results:
\begin{enumerate}
\item LOS polynomial foreground subtraction is non-optimal, and gives rise to larger error bars in final power spectrum than inverse-variance subtraction does.  This is mostly due to the fact that the LOS method projects out low-order polynomial modes, destroying information.  The inverse variance method, on the other hand, preserves all modes, even if it does downweights some of them substantially.  As a result, the error bars on the final power spectrum are larger for the LOS method (bottom panel of Figure \ref{tempuniterrors}) than for the inverse variance method (middle panel of Figure \ref{tempuniterrors}).
\item  {Traditional LOS polynomial subtraction methods contain residual noise and foreground biases in estimates of the power spectrum.}  Fortunately, this bias can be easily quantified and removed using Equation \ref{bias}.
\item In the low $k_\parallel$ regions of the $k_\perp$-$k_\parallel$ plane, the LOS polynomial foreground cleaning gives rise to power spectrum estimates that are highly correlated between neighboring $k_\parallel$ bins, or equivalently, to window functions that are elongated in the $k_\parallel$ direction.  The LOS polynomial subtraction has no effect on the $k_\perp$ direction and exhibits the same qualitative behavior regardless of the value of $k_\perp$ since it does not use angular foreground information in its cleaning.
\item The behavior of the inverse variance window functions, on the other hand, depends strongly on the value of $k_\perp$:
\begin{enumerate}
\item In regions where neither $k_\perp$ nor $k_\parallel$ is small, the post-inverse variance subtraction power spectrum estimates are no more correlated than what would be expected from the finiteness of the survey geometry alone. In other words, the inverse variance window functions in this region are unchanged by the presence of foregrounds.  This is due to the fact that foreground contamination was mild to begin with, resulting in a less aggressive foreground subtraction by the inverse variance weighting in order to keep power spectrum estimates as uncorrelated as possible.
\item In regions where both $k_\perp$ and $k_\parallel$ are small, the inverse variance window functions widen in both directions, but the effect in the $k_\perp$ direction is minimal.  This indicates that inverse variance subtraction operates mainly in the radial direction, even though this is not an  {a priori} mathematical constraint.  Put another way, the nature of the foregrounds and the availability of high spectral resolution instruments makes it much more fruitful to leverage frequency information (rather than angular information) in performing foreground subtraction.  This confirms intuitive expectations in the existing literature.
\item The inverse variance window functions show less $k_\parallel$ widening than the LOS polynomial window functions do, regardless of the location on the $k_\perp$-$k_\parallel$ plane.  This suggests that with the inverse variance scheme, one should be able to push to lower values of $k_\parallel$ before correlations between neighboring $k_\parallel$-bins make the resulting power spectrum estimates scientifically unusable.
\end{enumerate}
\end{enumerate}
In summary, the inverse-variance method that we have presented has several advantages over the LOS  polynomial subtraction methods proposed in the literature: they produce power spectrum measurements with smaller and less correlated error bars as well as narrower window functions.

\subsection{Applicability of Results to general arrays}
\label{genapp}
While the results presented in Section \ref{results} were computed for a set of specific MWA-like array parameters and a specific foreground model, we emphasize that the formalism presented in Section \ref{formalism} is completely general and can be applied to any $21\,\textrm{cm}$ tomography experiment and any foreground model.  In addition, as we will now argue, our qualitative conclusions from the last  subsection should in fact hold quite generally.

Consider first an experiment's location, antenna layout, integration time, rotation synthesis, and thermal noise level.  These factors may initially seem to form a high-dimensional parameter space that needs to be thoroughly explored, but from Section \ref{formalism} we know that the only place where they enter our mathematical framework is Equation \ref{noisecovar}, which determined our noise covariance matrix $\mathbf{N}$.  Their effects are thus degenerate with each other, and we saw in Section \ref{toy} that the resulting differences in the power spectrum estimation can be captured by a single parameter: the foreground-to-noise ratio $\gamma$.

Changing the frequency range of one's data cube has several effects, one of which is to alter the instrumental noise of the array, depending on the chosen frequency binning of the data.  This, too, is captured by the foreground-to-noise parameter.  More subtly, an increased frequency range changes the polynomial fit in the LOS scheme and affects foreground subtraction.  While this can result in large changes to the final power spectrum, it was shown in \citep{paper1} that for LOS subtraction to be effective in the first place, one \emph{must} perform the fits only over narrow frequency ranges anyway.  Moreover, as we have discussed above, cosmological evolution also imposes a constraint on the frequency range of any one data cube (one may of course separately analyze as many data cubes as necessary to cover the full data set).  Thus, the small frequency range used in Section \ref{results} is a good range to use, regardless of the full bandwidth of one's instrument.  For the inverse variance method, our conclusions are even more general than for the LOS method.  As we saw in the previous section, inverse variance subtraction can be thought of as a multiplicative filter in $k_\parallel$ space, and increasing the frequency range simply increases the resolution in $k_\parallel$-space without affecting the overall shape of the filter.

Another tunable parameter in our analysis is the flux cut $S_{cut}$ above which the bright point sources can be considered identifiable and removable in an earlier stage of data analysis.  In principle, varying $S_{cut}$ can affect the results in a complicated way, but from \citep{paper1}, we know that the amplitude of foregrounds essentially scales with $S_{cut}$, since the foregrounds are dominated by the brightest sources in a population.  Therefore, $S_{cut}$ is simply yet another parameter that can be folded into the foreground-to-noise ratio $\gamma$.

\subsection{Future challenges}

In this paper, we have shown that inverse variance foreground subtraction is able to clean foregrounds more effectively than traditional LOS methods can, resulting in power spectrum estimates with smaller error bars.  Perhaps the only disadvantage of the inverse variance method is its computational cost, since it requires the inversion of an $n_{pix} \times n_{pix}$ matrix, where $n_{pix} \sim n_\perp^2 n_\parallel$ is the total number of pixels in one's data cube.  An LOS polynomial algorithm, on the other hand, only requires the inversion of an $n_\parallel \times n_\parallel$ matrix.  Since matrix inversion naively scales as $\mathcal{O} (n^3)$, the difference is substantial.

Fortunately, by making a number of reasonable approximations one can speed up the method considerably.  As long as the fields of view are relatively narrow, a series of basis changes and matrix prewhitening techniques allow the relevant computations to be performed using algorithms that scale as $\mathcal{O}(n_{pix} \log n_{pix})$, and such algorithms are now being tested numerically for $21\,\textrm{cm}$ tomography \citep{josh}.  Moreover, since the inverse variance method operates mostly in the radial direction, one can define an ``effective" LOS algorithm that closely mimics the behavior of the full inverse variance subtraction.  This was demonstrated in Section \ref{toy}, where we developed an example of such an algorithm and saw that the inverse variance scheme can be thought of as a high-pass filter in the frequency direction, with the detailed properties of the filter dependent on noise and foreground characteristics.

To further improve the quality of foreground subtraction, it will be necessary to construct better foreground models.  This is because inverse variance foreground cleaning is not blind, but instead depends on our empirical knowledge of foreground properties.  It is important to note, however, that this is an advantage and not a disadvantage of the inverse variance method --- whereas the traditional LOS methods (being blind) will not improve with better knowledge of the foregrounds, the inverse variance method is virtually guaranteed to get better as $21\,\textrm{cm}$ tomography experiments begin to take science-quality data that are able to put tight constraints on our foreground models.  Combining the inverse variance method with the improved measurements to be expected from the low frequency radio astronomy community should therefore bring us closer to fulfilling the potential of $21\,\textrm{cm}$ tomography to improve our understanding of the Epoch of Reionization, the Dark Ages, and fundamental physics.

\section*{Acknowledgments}
The authors would like to thank Joshua Dillon for providing the data for the construction of Figure \ref{cubes} and for many helpful conversations, and Judd Bowman, Miguel Morales, Leo Stein, Christopher Williams, and Matias Zaldarriaga for helpful discussions.

\bibliography{powerspecwithfg}

\end{document}